\pdfoutput=1
\documentclass[a4paper,british,citeautoscript,floatfix,pdftex,superscriptaddress,twoside,%
aps,prl,%
longbibliography,%
reprint,twocolumn,final
]{revtex4-1}%
\usepackage{amsfonts,amsmath,amssymb}
\usepackage[T1]{fontenc}
\usepackage{graphicx}%
\usepackage[utf8]{inputenc}
\usepackage{microtype}
\usepackage[obeyFinal,textsize=footnotesize]{todonotes}
\usepackage{xspace}
\usepackage{hyperref, hypernat}
\usepackage[todos]{CMImacros}

\graphicspath{{./figures/}} 
\hypersetup{citebordercolor=yellow,linkbordercolor=red,urlbordercolor=blue}

\makeatletter%
\renewcommand*{\@fnsymbol}[1]{\ensuremath{\ifcase#1\or \|\or *\or **\or \mathparagraph\or
      \mathsection\or \dagger\or \ddagger\or \dagger\dagger \or \ddagger\ddagger \else\@ctrerr\fi}}
\makeatother

\newcommand*{\edfref}[1]{\hyperref[#1]{Extended Data Figure~\ref{#1}}\xspace}
\newcommand*{\suppinf}{Supplementary Information\xspace}%
\newcommand*{\methods}{\hyperref[sec:methods]{\emph{Methods}}\xspace}%

\newcommand{\cfeldesy}{\affiliation{Center for Free-Electron Laser Science, Deutsches
      Elektronen-Synchrotron DESY, Notkestraße 85, 22607 Hamburg, Germany}}%
\newcommand{\uhhcui}{\affiliation{The Hamburg Center for Ultrafast Imaging, Universität Hamburg,
      Luruper Chaussee 149, 22761 Hamburg, Germany}}%
\newcommand{\uhhphys}{\affiliation{Department of Physics, Universität Hamburg, Luruper Chaussee 149,
      22761 Hamburg, Germany}}%
\newcommand{\aarhuschem}{\affiliation{Department of Chemistry, Aarhus University, Langelandsgade
      140, 8000 Aarhus C, Denmark}}%
\newcommand{\mbi}{\affiliation{Max Born Institute, Max-Born-Straße 2a, 12489 Berlin, Germany}}%
\newcommand{\jkemail}{\email[]{jochen.kuepper@cfel.de}}%
\newcommand{\aremail}{\email[]{arnaud.rouzee@mbi.de}}%
\newcommand{\cmiweb}{\homepage{https://www.controlled-molecule-imaging.org}}%

\begin{document}
\title{Molecular movie of ultrafast coherent rotational dynamics}
\author{Evangelos T.\ Karamatskos}\cfeldesy\uhhphys%
\author{Sebastian~Raabe}\mbi
\author{Terry~Mullins}\cfeldesy%
\author{Andrea~Trabattoni}\cfeldesy\uhhphys%
\author{Philipp~Stammer}\mbi
\author{Gildas~Goldsztejn}\mbi
\author{Rasmus~R.\ Johansen}\aarhuschem
\author{Karol Długołęcki}\cfeldesy%
\author{Henrik Stapelfeldt}\aarhuschem%
\author{Marc~J.~J.\ Vrakking}\mbi%
\author{Sebastian Trippel}\cfeldesy\uhhcui%
\author{Arnaud Rouzée}\aremail\mbi%
\author{Jochen Küpper}\jkemail\cmiweb\cfeldesy\uhhphys\uhhcui%
\begin{abstract}\noindent%
   %
   Recording molecular movies on ultrafast timescales has been a longstanding goal for unravelling
   detailed information about molecular dynamics. We present the direct experimental recording of
   very-high-resolution and -fidelity molecular movies over more than one-and-a-half periods of the
   laser-induced rotational dynamics of carbonylsulfide (OCS) molecules. Utilising the combination
   of single quantum-state selection and an optimised two-pulse sequence to create a tailored
   rotational wavepacket, an unprecedented degree of field-free alignment, $\cost=0.96$
   ($\costhreeD=0.94$) was achieved, exceeding the theoretical limit for single-pulse alignment. The
   very rich experimentally observed quantum dynamics is fully recovered by the angular probability
   distribution obtained from solutions of the time-dependent Schrödinger equation with parameters
   refined against the experiment. The populations and phases of rotational states in the retrieved
   time-dependent three-dimensional wavepacket rationalised the observed very high degree of
   alignment.
\end{abstract}
\maketitle

The filming of nuclear motion during molecular dynamics at relevant timescales, dubbed the
``molecular movie'', has been a longstanding dream in the molecular
sciences~\cite{Zewail:JPCA104:5660, Ischenko:CR117:11066}. Recent experimental advances with
x-ray-free-electron lasers and ultrashort-pulse electron guns have provided first glimpses of
intrinsic molecular structures~\cite{Ayyer:Nature530:202, Kuepper:PRL112:083002,
   Hensley:PRL109:133202} and dynamics~\cite{Pande:Science352:725, Yang:PRL117:153002,
   Ischenko:CR117:11066}. However, despite the spectacular progress, the fidelity of the recorded
movies, in comparison to the investigated dynamics, was limited so far. Especially for
high-precision studies of small molecules, typically only distances between a few atoms were
determined~\cite{Kuepper:PRL112:083002, Hensley:PRL109:133202, Yang:PRL117:153002}.

Rotational quantum dynamics of isolated molecules provides an interesting and important testbed that
provides and requires direct access to angular coordinates. Furthermore, different from most
molecular processes, it can be practically exactly described by current numerical methods, even for
complex molecules. Rotational wavepackets were produced through the interaction of the molecule with
short laser pulses~\cite{Felker:JPC90:724, RoscaPruna:PRL87:153902, Stapelfeldt:RMP75:543}, which
couple different rotational states through stimulated Raman transitions. The resulting dynamics were
observed, for instance, by time-delayed Coulomb-explosion ion imaging~\cite{RoscaPruna:PRL87:153902,
   Mizuse:SciAdv1:e1400185, Dooley:PRA68:023406}, photoelectron
imaging~\cite{Marceau:PRL119:083401}, or ultrafast electron diffraction~\cite{Yang:NatComm7:11232}.
The rotational wavepackets were exploited to connect the molecular and laboratory frames through
strong-field alignment~\cite{RoscaPruna:PRL87:153902, Stapelfeldt:RMP75:543} and mixed-field
orientation~\cite{Ghafur:NatPhys5:289, Trippel:PRL114:103003}, as well as for the determination of
molecular-structure information in rotational-coherence spectroscopy~\cite{Felker:JPC96:7844,
   Riehn:CP283:297}. Coherent rotational wavepacket manipulation using multiple
pulses~\cite{Lee:PRL93:233601} or appropriate turn-on and -off timing~\cite{Trippel:PRA89:051401R}
allowed enhanced or diminished rephasing, and it was suggested as a realisation of quantum
computing~\cite{Lee:PRL93:233601}. Furthermore, methods for rotational-wavepacket reconstruction of
linear molecules~\cite{Mouritzen:JCP124:244311} and for benzene~\cite{Hasegawa:PRL101:053002} were
reported.

Here, we demonstrate the direct experimental high-resolution imaging of the time-dependent angular
probability-density distribution of a rotational wavepacket and its reconstruction in terms of the
populations and phases of field-free rotor states. Utilising a state-selected molecular sample and
an optimised two-laser-pulse sequence, a broad phase-locked rotational wavepacket was created. Using
mid-infrared-laser strong-field ionisation and Coulomb-explosion ion imaging, an unprecedented
degree of field-free alignment of $\cost=0.96$, or $\costhreeD=0.94$, was obtained at the full
revivals, whereas in between a rich angular dynamics was observed with very high resolution, from
which the complete wavepacket could be uniquely reconstructed. While the dynamics has low
dimensionality, The resulting --- purely experimentally obtained --- movie provides a most direct
realisation of the envisioned ``molecular movie''. We point out that the data also is a measurement
of a complete quantum carpet~\cite{Berry:PhysTod14:39}.

In order to achieve such a high degree of alignment, better than the theoretical maximum of
$\costhreeD=0.92$ for single-pulse alignment~\cite{Leibscher:PRL90:213001, Guerin:PRA77:041404}, we
performed a pump-probe experiment with ground-state-selected OCS molecules~\cite{Chang:IRPC34:557},
with $>80$~\% purity, as a showcase. Two off-resonant near-IR pump pulses of $800$~nm central
wavelength, separated by $38.1\,(1)$~ps and with a pulse duration of $250$~fs, \ie, much shorter
than the rotational period of OCS of $82.2$~ps, were used to create the rotational wavepacket. These
pulses were linearly polarised parallel to the detector plane. The probe pulse with a central
wavelength of $1.75~\um$ was polarised perpendicularly to the detector plane to minimise the effects
of geometric alignment and ensures that the observed degree of alignment was a lower boundary of the
real value. The probe pulse multiply ionised the molecules, resulting in Coulomb explosion into
ionic fragments. 2D ion-momentum distributions of O$^{+}$ fragments, which reflect the orientation
of the molecules in space at the instance of ionisation, were recorded by a velocity map imaging
(VMI) spectrometer~\cite{Eppink:RSI68:3477} for different time delays between the alignment pulse
sequence and the probe pulse. Further details of the experimental setup are presented in \methods.

\begin{figure}
   \includegraphics[width=\linewidth]{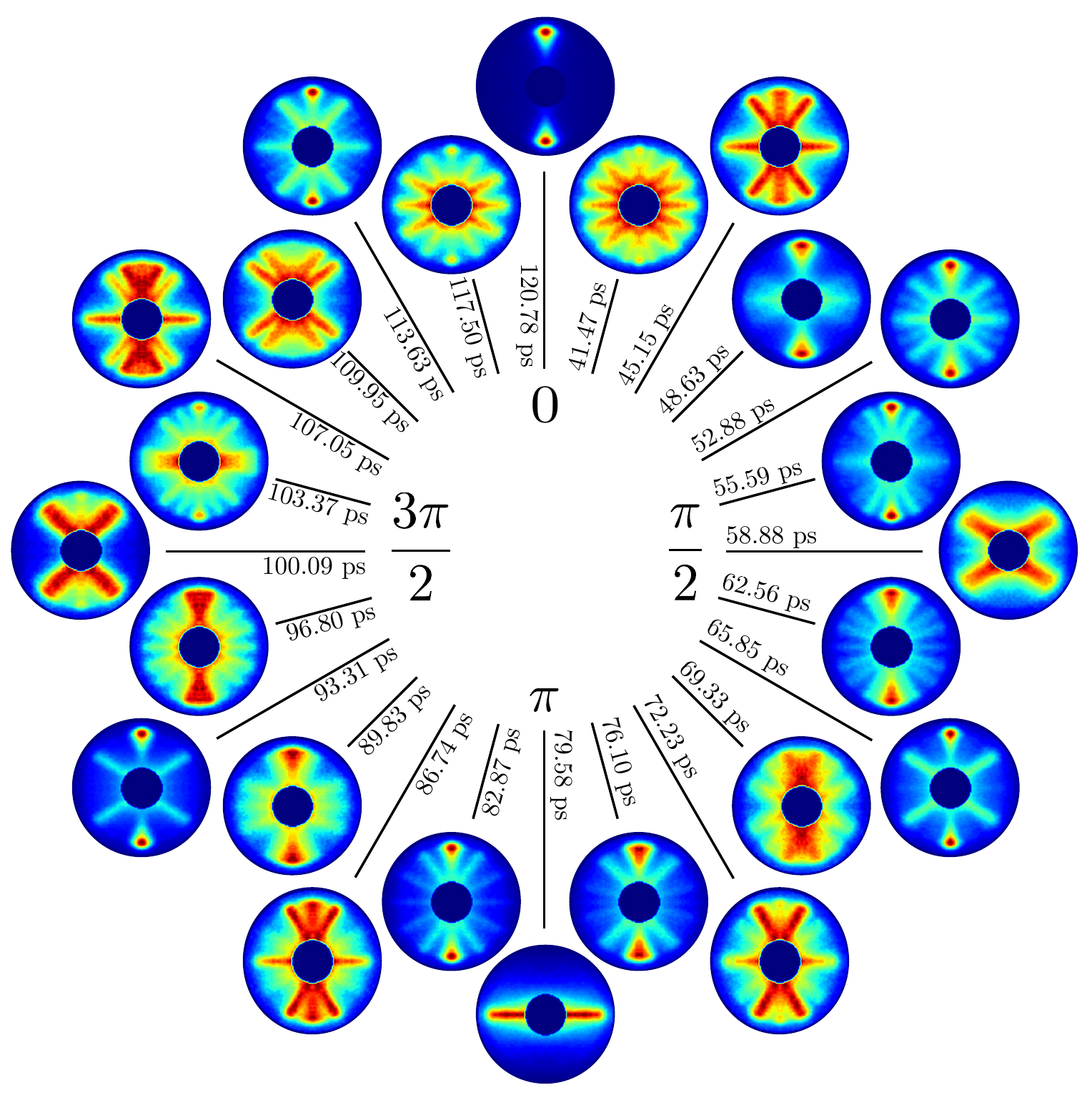}
   \caption{\textbf{Rotational clock depicting the molecular movie of the observed quantum
         dynamics.} Individual experimental VMI images of O$^+$ ion-momentum distributions depicting
      snapshots of the rotational wavepacket over one full period. The displayed data was recorded
      from the first (prompt) revival at 38.57~ps (0) to the first full revival at 120.78~ps
      ($2\pi$); the phase-evolution of $\pi/12$ between images corresponds to $\ordsim3.43$~ps and
      the exact delay times of the individual images are specified. Full movies are available as
      part of the \suppinf.}
   \label{fig:snapshots}
\end{figure}
In \autoref{fig:snapshots} snapshots of the experimentally recorded molecular movie, \ie, 2D
ion-momentum distributions, are shown for several probe times covering a whole rotational period.
The phase of $0$ and $2\pi$ correspond to $t=38.57$~ps and $120.78$~ps after the peak of the first
alignment laser pulse at $t=0$, respectively. The simplest snapshot-images, reflecting an
unprecedented degree of field-free alignment $\cost=0.96$, were obtained for the alignment revivals
at phases of $0$ and $2\pi$, which correspond to the prompt alignment and its revival regarding the
second laser pulse. Here, the molecular axes are preferentially aligned along the alignment-laser
polarisation. For the anti-alignment at a phase of $\pi$ the molecules are preferentially aligned in
a plane perpendicular to the alignment laser polarisation direction. Simple quadrupolar structures
are observed at $\pi/2$ and $3\pi/2$. At intermediate times, \eg, at $\pi/3$ or $7\pi/12$, the
images display rich angular structures, which could be observed due to the high angular experimental
resolution of the recorded movie, which is \degree{4} as derived in the \suppinf. This rich
structure directly reflects the strongly quantum-state selected initial sample exploited in these
measurements, whereas the structure would be largely lost in the summation of wavepackets from even
a few initially populated states.

The dynamics was analysed as follows: Through the interaction of the molecular ensemble with the
alignment laser pulses, a coherent wavepacket was created from each of the initially populated
rotational states. These wavepackets were expressed as a coherent superposition of eigenfunctions of
the field-free rotational Hamiltonian, \ie,
\begin{equation}
   \label{wavepacket}
   \Psi(\theta,\phi,t) = \sum_{J} a_{J}(t)Y_{J}^{M}(\theta,\phi) ,
\end{equation}
with the time-dependent complex amplitudes $a_{J}(t)$, the spherical harmonics
$Y_{J}^{M}(\theta,\phi)$, the quantum number of angular momentum $J$, and its projection $M$ onto
the laboratory-fixed axis defined by the laser polarisation. We note that $M$ was conserved and thus
no $\phi$ dependence existed. The angular distribution is defined as the sum of the squared moduli
of all $\Psi(\theta,\phi,t)$ weighted by the initial-state populations.

\begin{figure}
   \includegraphics[width=\linewidth]{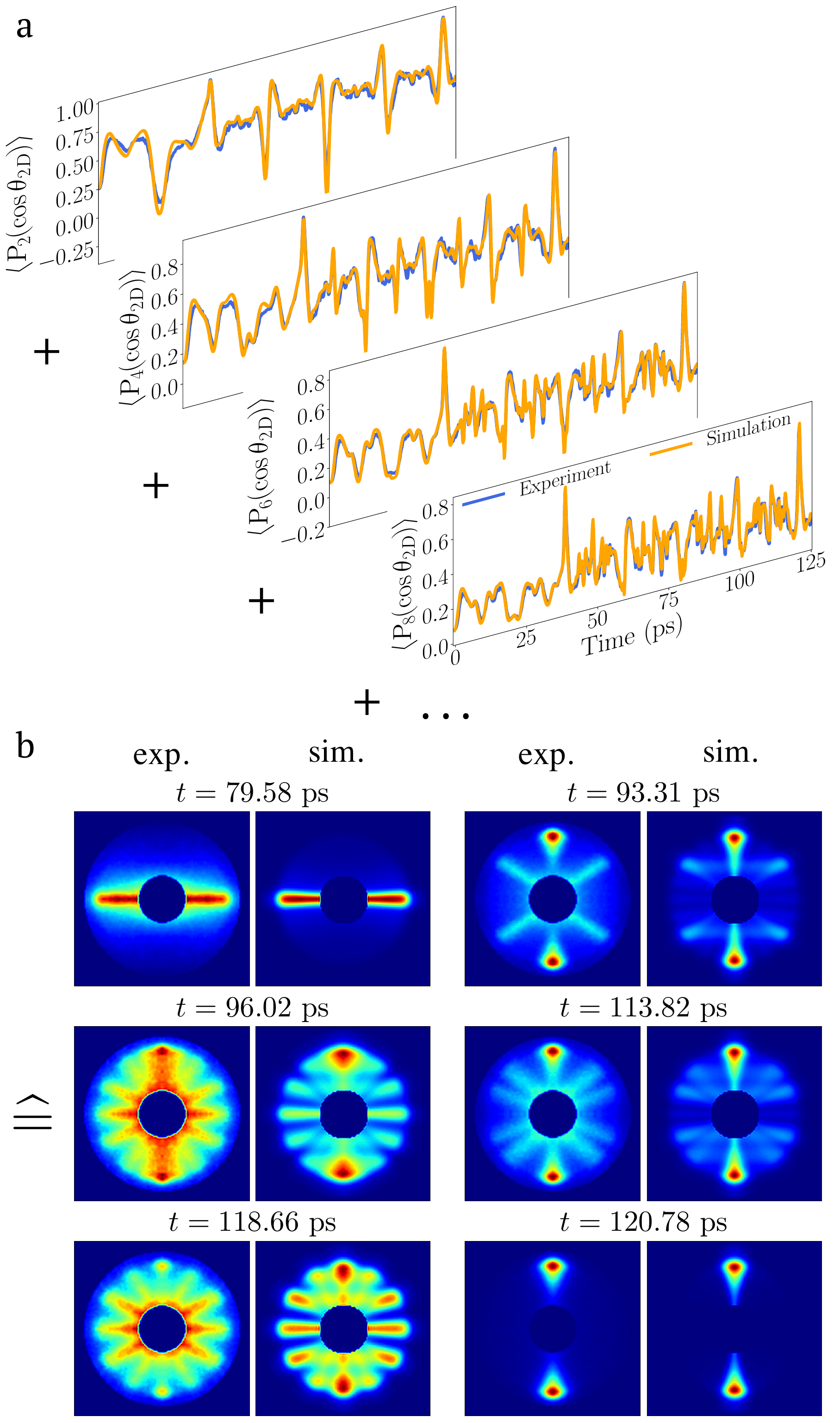}
   \caption{\textbf{Decomposition of angular distributions into their moments.} \quad \textbf{a}
      Comparison of the decomposition of the experimental and theoretical angular distributions in
      terms of Legendre polynomials. \textbf{b} Simulated and experimental angular-distribution VMI
      images for selected times; the radial distributions in the simulations are extracted from the
      experimental distribution at 120.78~ps, see text for details.}
   \label{fig:moments}
\end{figure}
The degree of alignment was extracted from the VMI images using the commonly utilised expectation
value $\cost$. The maximum value observed at the alignment revival reached $0.96$, which, to the
best of our knowledge, is the highest degree of field-free alignment achieved to date. Comparing the
angular distributions at different delay times with the degree of alignment $\cost$, see Fig.~S4 in
the \suppinf, we observed the same degree of alignment for angular distributions that are in fact
very different from each other. This highlights that much more information is contained in the
angular distributions than in the commonly utilised expectation value~\cite{Stapelfeldt:RMP75:543}.
Indeed, \cost merely describes the leading term in an expansion of the angular distribution, for
instance, in terms of Legendre polynomials, see (1) in the \suppinf. In order to fully characterise
the angular distribution a description in terms of a polynomial series is necessary that involves
the same maximum order as the maximum angular momentum $J_\text{max}$ of the populated rotational
eigenstates, which corresponds to, at most, $2J_\text{max}$ lobes in the momentum maps.

As the probe laser is polarised perpendicularly to the detector plane, the cylindrical symmetry as
generated by the alignment-laser polarisation was broken and an Abel inversion to retrieve the 3D
angular distribution directly from the experimental VMI images was not possible. In order to
retrieve the complete 3D wavepacket, the time-dependent Schrödinger equation (TDSE) was solved for a
rigid rotor coupled to a non-resonant ac electric field representing the two laser pulses as well as
the dc electric field of the VMI spectrometer. For a direct comparison with the experimental data
the rotational wavepacket and thus the 3D angular distribution was constructed and, using a
Monte-Carlo approach, projected onto a 2D screen using the radial distribution extracted from the
experiment at the alignment revival at 120.78~ps. The relation between the 3D rotational wavepacket
and the 2D projected density is graphically illustrated in Fig.~S2 of the \suppinf. The anisotropic
angle-dependent ionisation efficiency for double ionisation, resulting in a two-body breakup into
O$^{+}$ and CS$^{+}$ fragments, was taken into account by approximating it by the square of the
measured single-electron ionisation rate. Non-axial recoil during the fragmentation process is
expected to be negligible and was not included in the simulations.

The initial state distribution in the quantum-state selected OCS sample as well as the interaction
volume with the alignment and probe lasers were not known \emph{a priori} and used as fitting
parameters. For each set of parameters the TDSE was solved and the 2D projection of the rotational
density, averaged over the initial state distribution and the interaction volume of the pump and
probe lasers, was carried out. The aforementioned expansion in terms of Legendre polynomials was
realised for the experimental and simulated angular distributions and the best fit was determined
through least squares minimisation, see \suppinf. Taking into account the eight lowest even moments
of the angular distribution allowed to precisely reproduce the experimental angular distribution.
The results for the first four moments are shown in \autoref[a]{fig:moments}; the full set is given
in Fig.~S3 in the \suppinf as well as the optimal fitting parameters. The overall agreement between
experiment and theory is excellent for all moments. Before the onset of the second pulse, centred
around $t=38.1$~ps, the oscillatory structure in all moments is fairly slow compared to later times,
which reflects the correspondingly small number of interfering states in the wavepacket before the
second pulse, and the large number thereafter.

Theoretical images, computed for the best fit parameters, are shown in \autoref[b]{fig:moments}; a
full movie is provided with the \suppinf. The theoretical results are in excellent agreement with
the measured ion-momentum angular distributions at all times, see \suppinf, and prove that we were
able to fully reconstruct the 3D rotational wavepacket with the amplitudes and phases of all
rotational states included.

\begin{figure}
   \includegraphics[width=\linewidth]{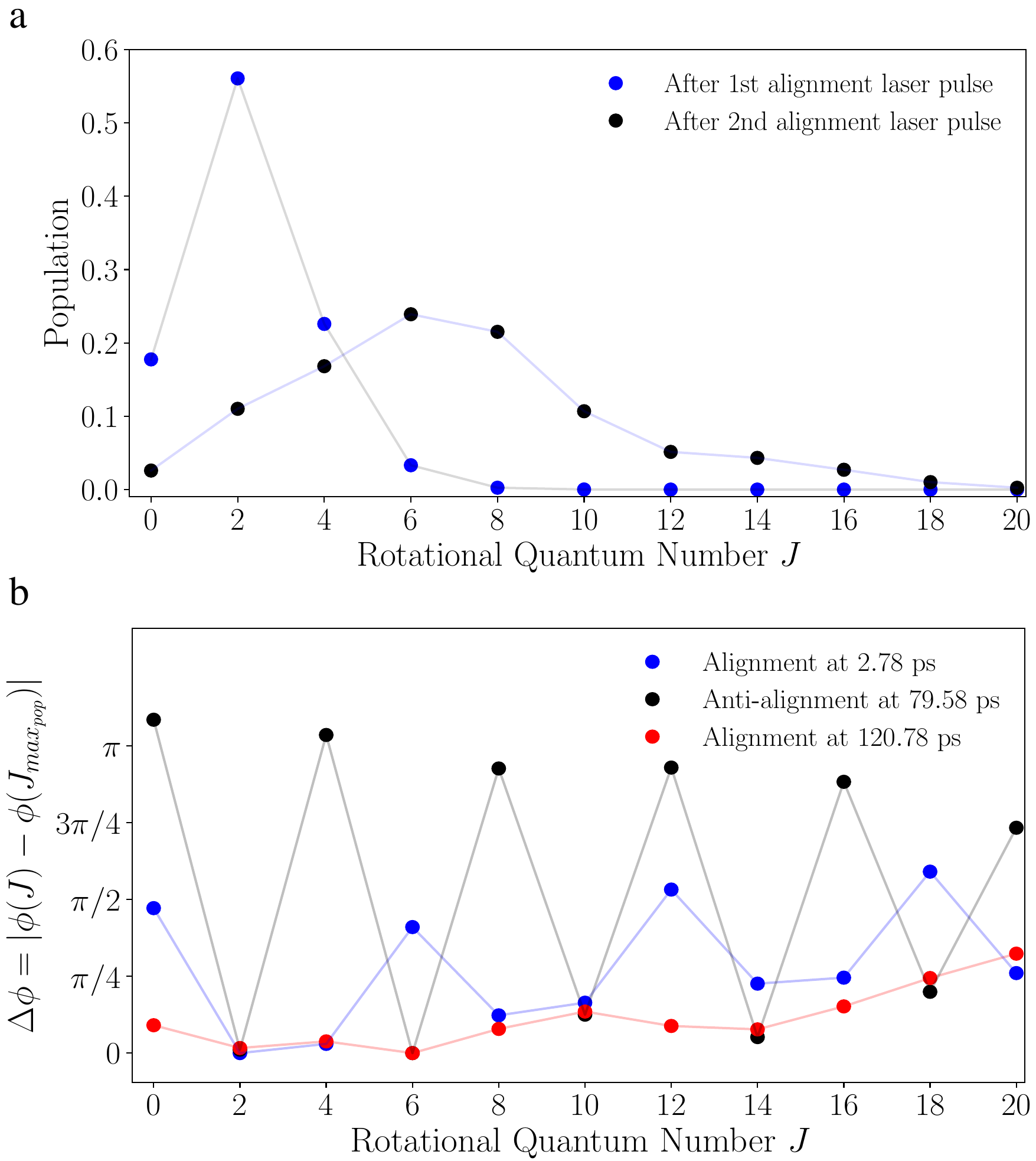}
   \caption{\textbf{Populations and phase differences in the rotational wavepacket at alignment and
         anti-alignment times} \newline \textbf{a} Rotational-state populations and \textbf{b}
      phase-differences to the phase of the state with largest population, $J=2,J=6$, respectively,
      at the alignment revival following a single-pulse excitation, $2.78$~ps (blue dots), and the
      two-pulse excitation, $120.78$~ps (red dots) as well as for the antialignment at $79.58$~ps
      (black dots, populations coincide with the red dots). Only states with even angular momentum
      are populated due to the Raman-transition selection rules $\Delta{J}=\pm2$.}
   \label{fig:pop+phase}
\end{figure}
In \autoref[a]{fig:pop+phase}, the extracted rotational-state populations are shown for the
wavepacket created from the rotational ground state after the first and the second alignment laser
pulse. It clearly shows that the rotational-state distribution is broader after the second pulse,
reaching up to $J\ge16$. This also matches the convergence of the Legendre-polynomial series, with
eight even terms, derived from the fit to the data above. In \autoref[b]{fig:pop+phase} the
corresponding phase differences for all populated states relative to the initial state with the
largest population in the wavepacket are shown, where $\phi(J)$ is the phase of the complex
coefficient $a_{J}$ in \eqref{wavepacket}. Combining these populations and phases it became clear
that the very high degree of alignment after the second alignment pulse arises from the combination
of the broad distribution of rotational states, reaching large angular momenta, and the very strong
and flat rephasing of all significantly populated states at the revival at $120.78$~ps, \autoref[b
(red)]{fig:pop+phase}. Similarly, the anti-alignment at $79.58$~ps occurs due to alternating phase
differences of $\pi$ between adjacent populated rotational states, \autoref[b
(black)]{fig:pop+phase}.

In conclusion, we were able to record a high-resolution molecular movie of the ultrafast coherent
rotational motion of impulsively aligned OCS molecules. State-selection and an optimised two-pulse
sequence yielded an unprecedented degree of field-free alignment of $\cost=0.96$, with a very narrow
angular confinement of $\degree{13.4}$ FWHM. Limiting the analysis to a determination of \cost, as
it is common in experiments on time-dependent alignment, did not allow to capture the rich
rotational dynamics, while the use of a polynomial expansion up to an appropriate order did. We
reconstructed the rotational wavepacket, from which the complex coefficients and, hence, the full
information about the rotational wavepacket under study was extracted. The 2D projection of the
reconstructed rotational wavepacket allowed a direct comparison with the experimentally measured
data.

Regarding the extension toward the investigation of chemical dynamics, we point out that
strong-field-ionisation-induced Coulomb-explosion imaging can be used, for instance, to image the
configuration of chiral molecules~\cite{Pitzer:Science341:1096} or internal torsional
dynamics~\cite{Christensen:PRL113:073005}. Following the dynamics of such processes with the detail
and quality presented here would directly yield a molecular movie of the chemical and, possibly,
chirality dynamics~\cite{Owens:PRL121:193201}. Furthermore, the very high degree of field-free
alignment achieved here would be extremely useful for stereochemistry
studies~\cite{Kuipers:Nature334:420, Rakitzis:Science303:1852} as well as for molecular-frame
imaging experiments~\cite{Itatani:Nature432:867, Holmegaard:NatPhys6:428, Filsinger:PCCP13:2076,
   Hensley:PRL109:133202, Kuepper:PRL112:083002, Yang:NatComm7:11232, Weber:PRL111:263601,
   Pullen:NatComm6:7262, Barty:ARPC64:415}.

\section{Methods}
\label{sec:methods}
A cold molecular beam was formed by supersonic expansion of a mixture of OCS ($500$~ppm) in helium,
maintained at a backing pressure of $90$~bar from a pulsed Even-Lavie
valve~\cite{Hillenkamp:JCP118:8699} operated at $250$~Hz. After passing two skimmers, the collimated
molecular beam entered the Stark deflector. The beam was dispersed according to quantum state by a
strong inhomogeneous electric field~\cite{Chang:IRPC34:557} with a nominal strength of
$\ordsim200$~kV/cm. Through a movable third skimmer, the molecular beam entered the spectrometer.
Here, it was crossed at right angle by laser beams, where the height of the laser beams allowed to
probe state-selected molecular ensembles, \ie, a practically pure rovibronic-ground-state sample of
OCS~\cite{Nielsen:PCCP13:18971, Trippel:PRA89:051401R, Trippel:PRL114:103003}.

The laser setup consisted of a commercial Ti:Sapphire laser system (KM labs) delivering pulses with
$30$~mJ pulse energy, $35$~fs (FWHM) pulse duration, and a central wavelength of 800~nm at a $1$~kHz
repetition rate. One part (20~mJ) of the laser output was used to pump a high-energy tunable optical
parametric amplifier (HE-TOPAS, Light Conversion) to generate pulses with a central wavelengths of
$1.75~\um$, a pulse duration of 60~fs, and a pulse energy of $\ordsim1.5$~mJ. $900~\uJ$ of the
remaining $800$~nm laser output was used for the laser-induced alignment, \ie, the generation of the
investigated rotational wavepackets. This beam was split into two parts with a 4:1 energy ratio
using a Mach-Zehnder interferometer. A motorised delay stage in one beam path allowed for
controlling the delay between the two pulses. This delay was optimised experimentally and maximum
alignment was observed for $\tau_{exp}=38.1\pm0.1$~ps, in perfect agreement with the theoretically
predicted $\tau_{sim}=38.2$~ps. The pulses were combined collinearly and passed through a 2~cm long
$\text{SF}_{11}$ optical glass to stretch them to $250$~fs pulse duration (FWHM). Then the alignment
pulses were collinearly overlapped with the 1.75~\um mid-infrared pulses using a dichroic mirror.
All pulses were focused into the velocity map imaging spectrometer (VMI) using a 25~cm
focal-distance calcium fluoride lens.

At the center of the VMI the state-selected molecular beam and the laser beams crossed at right
angle. Following strong-field multiple ionisation of the molecules, the generated charged fragments
were projected by the VMI onto a combined multichannel-plate (MCP) phosphor-screen detector and read
out by a CCD camera. The angular resolution of the imaging system is \degree{4}, limited by the
1~megapixel camera, see \suppinf. 2D ion-momentum distributions of O$^+$ fragments were recorded as
a function of the delay between the $800$~nm pulses and the ionising $1.75~\um$ pulses in order to
characterise the angular distribution of the molecules through Coulomb-explosion imaging. The
polarisation of the $800$~nm alignment pulses was parallel to the detector screen whereas that of
the $1.75~\um$ ionising laser was perpendicular in order to avoid geometric-alignment effects in the
angular distributions. For this geometry, unfortunately, it was not possible to retrieve 3D
distributions from an inverse Abel transform. 651 images were recorded in steps of $193.4$~fs,
covering the time interval from -0.7~ps up to 125~ps, which is more than one and a half times the
rotational period of OCS of 82.2~ps.

\section{Optimisation of two-pulse field-free alignment}
Optimisation calculations were performed in order to predict the optimal pulse parameter for single
and two-pulse field-free alignment. In the simulations, the rotational part of the Schrödinger
equation for a linear rigid rotor within the Born-Oppenheimer approximation coupled to non-resonant
ac alignment laser pulses and a static electric field, as provided by the VMI in the interaction
region, was used. The Hamiltonian of the system is described in detail in
reference~\cite{Omiste:PRA86:043437}. The global differential-evolution-optimisation
algorithm~\cite{Storn:JGP11:341} was used to calculate the optimal alignment characterised through
the expectation value $\costhreeD$ in a closed-feedback-loop approach. The optimisation parameters
used were the intensities and one common duration of Fourier-limited Gaussian pulses and the delay
between the pulses in the case of two-pulse alignment. In the calculations a pure rotational ground
state ensemble was assumed and no integration over the interaction volume was carried out. The
former is justified as we know that the ground state contribution to alignment is dominant and the
exact experimental conditions were not known a priori. Furthermore, exploiting the electrostatic
deflector, as in our experiment, almost pure ground state ensembles can be
prepared~\cite{Nielsen:PCCP13:18971, Chang:IRPC34:557}. Including also thermally excited rotational
states lead to an additional incoherent sum over all states present in the initial distribution of
rotational states and in general to a decrease of the degree of alignment. The same holds for the
interaction volume of the laser since only molecules at the center of the beam experience the
optimal alignment intensity while molecules at some distance from the center interact with a lower
field. In this sense the calculated values constitute an upper limit for the alignment under optimal
conditions.
\begin{figure}
   \includegraphics[width=\linewidth]{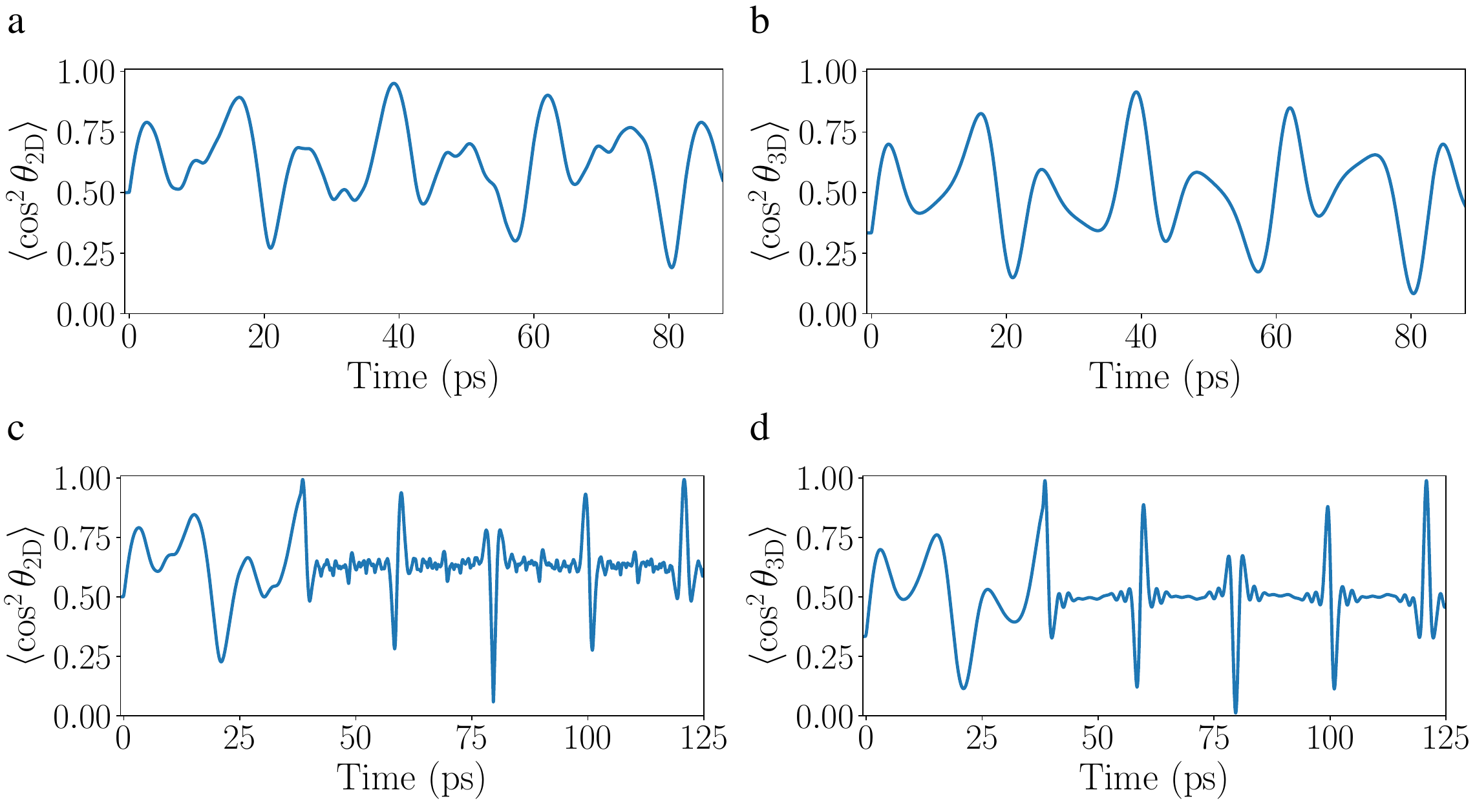}
   \caption{Optimised \textbf{a, c} 2D and \textbf{b, d} 3D field-free alignment with \textbf{a, b}
      one and \textbf{c, d} two alignment pulses.}
   \label{fig:alignment}
\end{figure}
Time-evolution curves of the optimal two-dimensional (2D) and three-dimensional (3D) alignment are
shown in~\autoref{fig:alignment}. The results for the single pulse optimisation yielded a pulse
duration of $114.5$~fs and a maximum intensity of $5.6$~TW/cm$^{2}$. The corresponding maximum
degree of field-free 3D alignment was found to be $\costhreeD=0.92$, which is in agreement with an
upper bound of $0.92$ derived previously~\cite{Guerin:PRA77:041404}; this corresponds to a 2D degree
of alignment of $\cost=0.95$.

The optimal parameters for the case of two alignment pulses were found to be a pulse duration of
$273.4$~fs, a pulse separation between the two pulses of $38.2$~ps, and an intensity-ratio of
$\ordsim1:5$ with the first pulse being weaker than the second one, in agreement with previous
results~\cite{Bisgaard:PRL92:173004, Leibscher:PRL90:213001}. The maximum intensity of the first
pulse was determined to be $1.93$~TW/cm$^{2}$ and that of the second pulse $10$~TW/cm$^{2}$, which
was the upper bound of intensities included in the calculations since for higher values at a
wavelength of $800$~nm a non-negligible amount of ionisation of OCS sets in. The maximum degree of
3D field-free alignment calculated with these parameters were $\costhreeD=0.99$ and $\cost=0.99$,
substantially higher than in the single pulse case. The experiment presented in the main paper was
performed under experimental conditions approximating these optimised parameter. We note that the
optimal pulse separation was calculated to be $38.2$~ps, which was confirmed in the experiment, for
which a scan of the pulse separation yielded the best alignment revival for $38.1\pm0.1$~ps.

\section{Moments of angular distribution}
There are several ways to expand the angular distribution of the wavepacket in a power
series~\footnote{We had indeed originally performed the analysis in terms of squared Chebyshev
   polynomials \expectation{\cos^{2}{n\theta_{2D}}} for numerical convenience and the results of
   both approaches are identical.}, but a natural basis consists of the Legendre polynomials, as for
$\Delta{M}=0$ the eigenstates are independent of $\phi$ and the spherical harmonics simplify to
Legendre polynomials. Only even order polynomials appear in the expansion since for a
ground-state-selected ensemble the odd order moments describe orientation of the molecular axes,
which was not present. The expansion takes on the form
\begin{equation}
   \label{eqn:expansion}
   P(\theta_{2D},t) = \sum_{k=0,k\,\mathrm{even}}^{J_{\text{max}}} a_{k}(t)\mathcal{P}_{k}(\cos{\theta_{2D}})
\end{equation}
where the full time-dependent angular distribution is denoted as $P(\theta_{2D},t)$ and $a_{k}$
\sloppy\mbox{($k=0,2\ldots{}J_\text{max}$)} are the expansion coefficients corresponding to the
$k$-th Legendre polynomial $\mathcal{P}_{k}$; $J_\text{max}$ is the angular momentum quantum number
of the highest populated rotational state in the wavepacket.

\begin{figure}[b]
   \includegraphics[width=\linewidth]{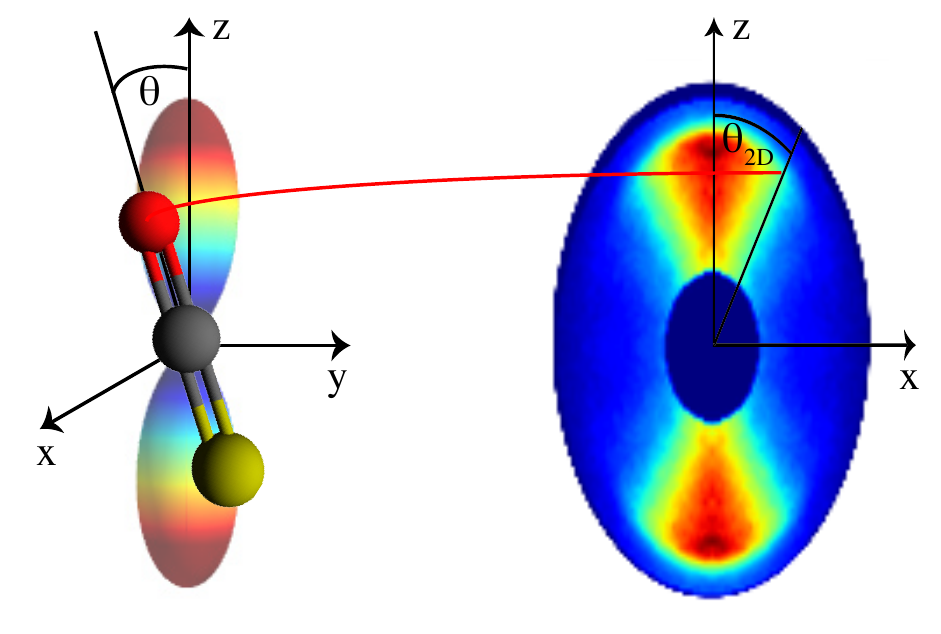}
   \caption{Relation between the Euler angle $\theta$, defining the alignment of the molecular axis
      with respect to the pump laser polarisation axis, and $\theta_{2\text{D}}$, corresponding to
      the angle between the pump laser polarization and the detected ion-momentum distribution on
      the 2D detector.}
   \label{fig:angles}
\end{figure}

In order to characterise the initial state distribution of rotational states in the molecular beam,
the eight lowest even-order moments of the experimental angular distributions were fitted
simultaneously using least squares minimisation. For each moment, squared differences were summed
according to
\begin{equation}
   \chi_{n}^{2} = \sum_{t} \left( \langle \mathcal{P}_{2n}(\cos{\theta_{2D}})_\text{exp}\rangle (t)
      - \langle \mathcal{P}_{2n}(\cos{\theta_{2D}})_\text{sim,vol}\rangle (t)
   \right)^2 \, ,
   \label{eqn:least}
\end{equation}
where the sum runs over all measured delay times $t$ and $n=1\ldots8$. In order to compute
$\langle\mathcal{P}_{n}(\cos{\theta_{2D}})_\text{sim,vol}\rangle(t)$, several steps were followed.
First, the coherent wavepackets, created through the interaction with the alignment laser pulses,
were for every initial state described in the basis of field-free eigenstates as
\begin{equation}
   \Psi_{J_\text{i}, M_\text{i}}(\theta,\phi,t) = \sum_J a_J(t) Y_J^{M_\text{i}}(\theta,\phi),
   \qquad a_J(t=0)=\delta_{JJ_\text{i}} ,
\end{equation}
where $a_{J}(t)=|a_{J}(t)|e^{i\phi_{J}(t)}$ are time-dependent complex coefficients with amplitude
$|a_{J}(t)|$, phase $\phi_{J}(t)$, and initial condition $a_{J}(t=0)=\delta_{JJ_{\text{i}}}$,
$\delta_{JJ_\text{i}}$ is the Kronecker delta, obtained from the solution of the time-dependent
Schrödinger equation; $Y_J^M(\theta,\phi)$ are the spherical-harmonic functions and
$J_\text{i}, M_\text{i}$ are the quantum numbers of the initial state from which the wavepacket is
formed. The sum runs only over $J$, since $M$ was a good quantum number due to cylindrical symmetry,
as imposed by the linear polarisation of the alignment laser, and, hence, $\Delta{M}=0$ and
$M=M_\text{i}$ was conserved. Furthermore, the selection rules for transitions between different
rotational states were $\Delta{J}=\pm2$, since the population transfer is achieved via non-resonant
two-photon Raman transitions. Moreover, the static VMI field was perpendicular to the alignment
laser polarisation and does not mix different $M$ states. Since more than one rotational state were
initially populated, the 3D rotational density was obtained through the incoherent average with
statistical weights $w_{J_\text{i},M_\text{i}}$
\begin{equation}
   P_{\text{sim,3D}}(\theta,\phi,t) = \sum_{J_{\text{i}}, M_{\text{i}}} w_{J_{\text{i}}, M_{\text{i}}}\, p(\theta)
   \left|\Psi_{J_{\text{i}}, M_{\text{i}}}(\theta,\phi,t)\right|^{2} \, ,
\end{equation}
which were not known \emph{a priori} and used as fitting parameters. The function $p(\theta)$
describes the angle-dependent ionisation probability, which was approximated through the square of
the measured angular-dependent single-electron ionisation rate. Finally, a focal average over the
interaction region with the alignment and probe laser beam profiles, assumed to be Gaussian, was
performed. The average over intensities in the laser focus was calculated through integration
\begin{multline}
   \label{volume_average}
   \expectation{P_{\text{sim,3D}}(\theta,\phi,t)}_\text{vol}(t) = \\ 
   = \frac{1}{N} \int_{0}^{r_\text{max}} \expectation{P_{\text{sim,3D}}(\theta,\phi,\Ialign(r),t)}
      e^{-2r^{2}/w^{2}_\text{probe}} \; r\mathrm{d}r
\end{multline}
with radius $r_\text{max}$ at $\Ialign=10^9$~W/cm$^2$ and $N$ a normalisation factor. The dependence
of the rotational wavepackets on the alignment laser intensities is explicitly stated in
\eqref{volume_average}. The widths of the laser beams were also not known \emph{a priori} and were
included as further fitting parameters. The resulting focal- and initial-state-averaged 3D
rotational densities were projected onto a 2D plane using a Monte-Carlo sampling routine, which
included the experimental radial distribution extracted at the full revival at a delay time of
$120.78~$ps, yielding the simulated VMI images in Fig.~2 in the main paper. The relation between the
3D rotational density and the 2D projected density is graphically illustrated in
\autoref{fig:angles}.
The Legendre moments of the angular distribution were then extracted from the 2D projected images
and compared to experiment through $\chi^{2}_{n}$, as described in \eqref{eqn:least}. The
statistical weights $w_{J_\text{i}M_\text{i}}$ of the initial state distribution and the laser focal
sizes were varied until $\sum_{n}\chi^{2}_{n}$ converged to its minimum. The individual populations
determined through the fitting procedure are $w_{00}=8.2\cdot 10^{-1}$, $w_{10}=3.7\cdot 10^{-2}$,
$w_{11}=7.5\cdot 10^{-2}$, $w_{20}=1.5\cdot 10^{-2}$, $w_{21}=2.1\cdot 10^{-2}$ and
$w_{22}=3.2\cdot 10^{-2}$ and the optimal focal parameter were determined to be
$w_\text{align}=130~\um$ for the alignment laser and $w_\text{probe}=60~\um$ for the probe laser.
The results are consistent with the fact that the probe laser was tighter focused than the alignment
laser such that only molecules exhibiting strong alignment, close to the beam center, are probed.

\begin{figure}
   \includegraphics[width=\linewidth]{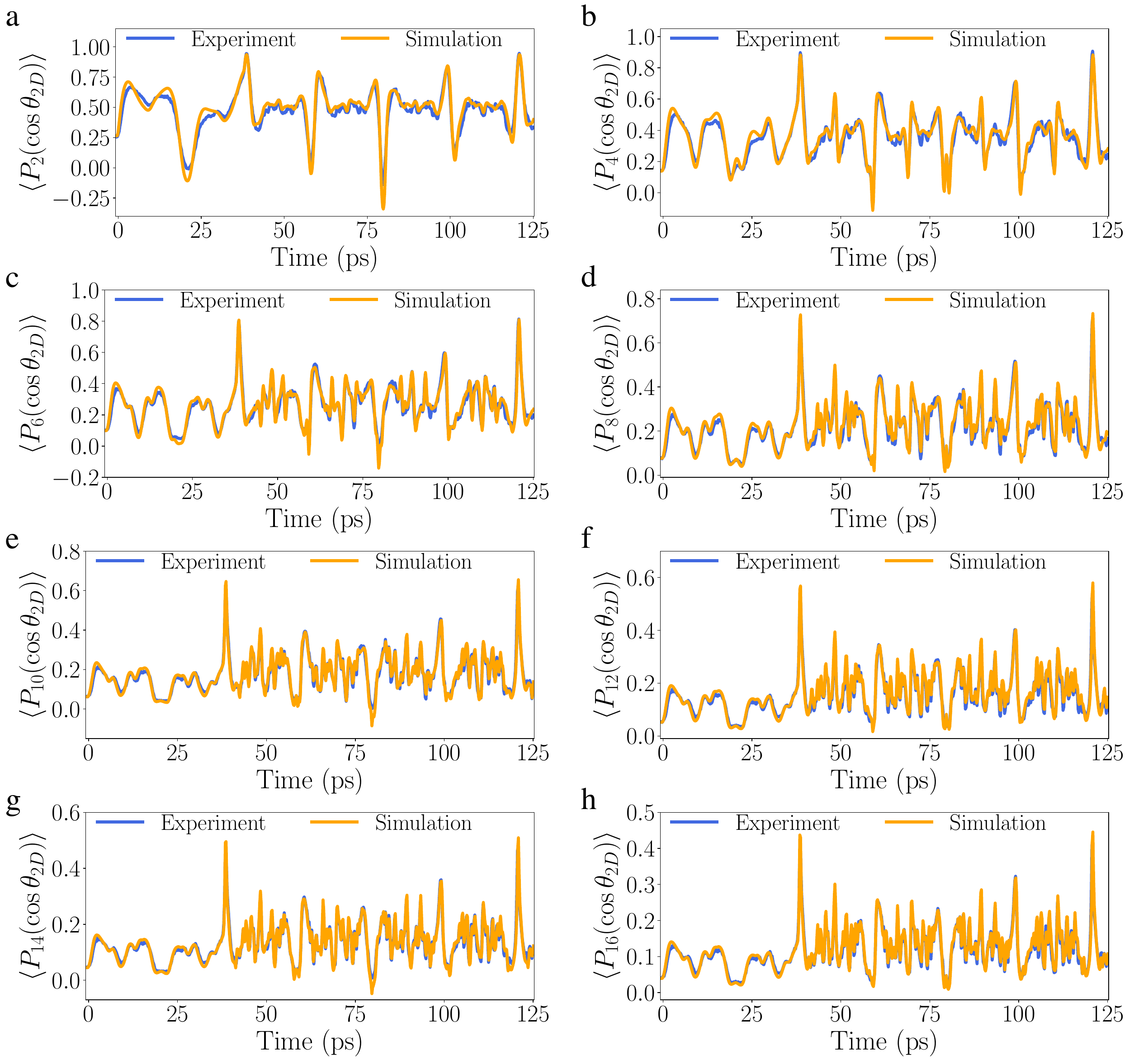}
   \caption{Even order moments 1 to 8 of the angular distribution \textbf{a}
      \expectation{P_{2}(\cos{\theta_{2D}})}, \textbf{b} \expectation{P_{4}(\cos{\theta_{2D}})},
      \textbf{c} \expectation{P_{6}(\cos{\theta_{2D}})}, \textbf{d}
      \expectation{P_{8}(\cos{\theta_{2D}})} \textbf{e} \expectation{P_{10}(\cos{\theta_{2D}})},
      \textbf{f} \expectation{P_{12}(\cos{\theta_{2D}})}, \textbf{g}
      \expectation{P_{14}(\cos{\theta_{2D}})}, \textbf{h} \expectation{P_{16}(\cos{\theta_{2D}})}.}
   \label{fig:moments}
\end{figure}
The final results of the fitting procedure are shown in Fig.~2 in the main paper and in
\autoref{fig:moments}. The simulated angular distributions and the moments of the angular
distribution are in excellent agreement with the experiment, in particular all oscillations are
correctly captured, even for the highest-order moments. The experimental parameters used for the
simulations were the peak intensities for the two alignment pulses of
$\Ialign{_{,1}}=1.92\cdot10^{12}$~W/cm$^{2}$ and $\Ialign{_{,2}}=5.5\cdot10^{12}$~W/cm$^{2}$, the
pulse duration of the alignment laser pulses $\tau_\text{align}=255$~fs, the time delay between the
two alignment laser pulses $\tau_\text{delay}=38.1$~ps, and the pulse duration of the probe laser
$\tau_\text{probe}=60$~fs. Calculations with 21 initial states, \ie, $J=0\ldots5,M=0\ldots5$,
included in the initial rotational state distribution were originally performed, but convergence was
already reached for the 6 lowest-energy states and the fitting procedure was restricted to using
these 6 lowest rotational states, \ie, $J=0\ldots2,M=0\ldots2$, and the focal volume was averaged
over 100 intensities in $\Ialign=1\cdot 10^9\ldots 5.5\cdot 10^{12}$~W/cm$^2$. In all calculations
the basis for each coherent wavepacket included all rotational states up to $J=50$.

\section{Angular distributions}
\begin{figure}
   \includegraphics[width=\linewidth]{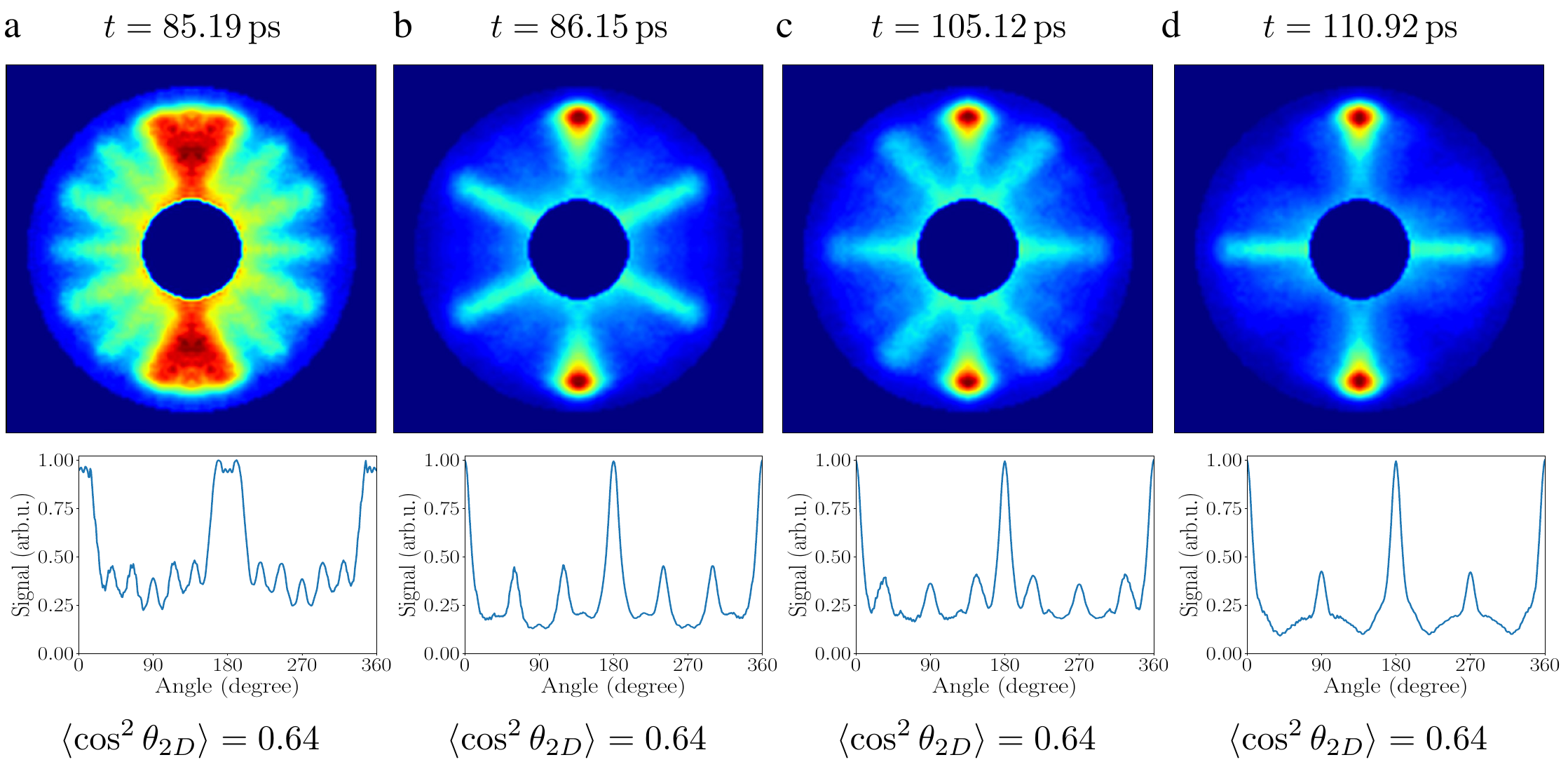}
   \caption{Four O$^{+}$ ion momentum probability distributions recorded at different time delays,
      displaying very different angular distributions, but having all the same degree of alignment
      of $\cost=0.64$. These distributions were recorded for delay times of \textbf{a} $85.19$~ps,
      \textbf{b} $86.15$~ps, \textbf{c} $105.12$~ps, and \textbf{d} $110.92$~ps.}
   \label{fig:ang_distributions}
\end{figure}
As pointed out in the main text, we observed angular probability distributions showing the very rich
time evolution of the rotational wavepacket created by the two alignment laser pulses. When
characterising the degree of alignment using the commonly used \cost we observed that completely
different angular distributions possess the same degree of alignment, which pointed out the need for
higher order terms in the expansion of the total angular distribution, \eg, in the basis of Legendre
polynomials, to be able to reconstruct the complete rotational wavepacket. In
\autoref{fig:ang_distributions} we present O$^{+}$ ion momentum distributions measured at four
different delay times together with their corresponding angular distributions corroborating this
observation. The delay times were chosen such that all distributions have the same $\cost=0.64$,
corresponding to the permanent alignment level. Although the degree of alignment is quite low
compared to the maximum degree of alignment achieved, one clearly sees in particular in
\autoref[b--d]{fig:ang_distributions} that nevertheless there is a substantial amount of molecules
being strongly aligned. Thus it is clearly not sufficient to just use the degree of alignment in
terms of \cost to characterise the molecular alignment distribution, but the knowledge of the whole
angular distribution is needed.

\subsection{Comparison of angular distributions from experiment and simulations}
\begin{figure}
   \includegraphics[width=\linewidth]{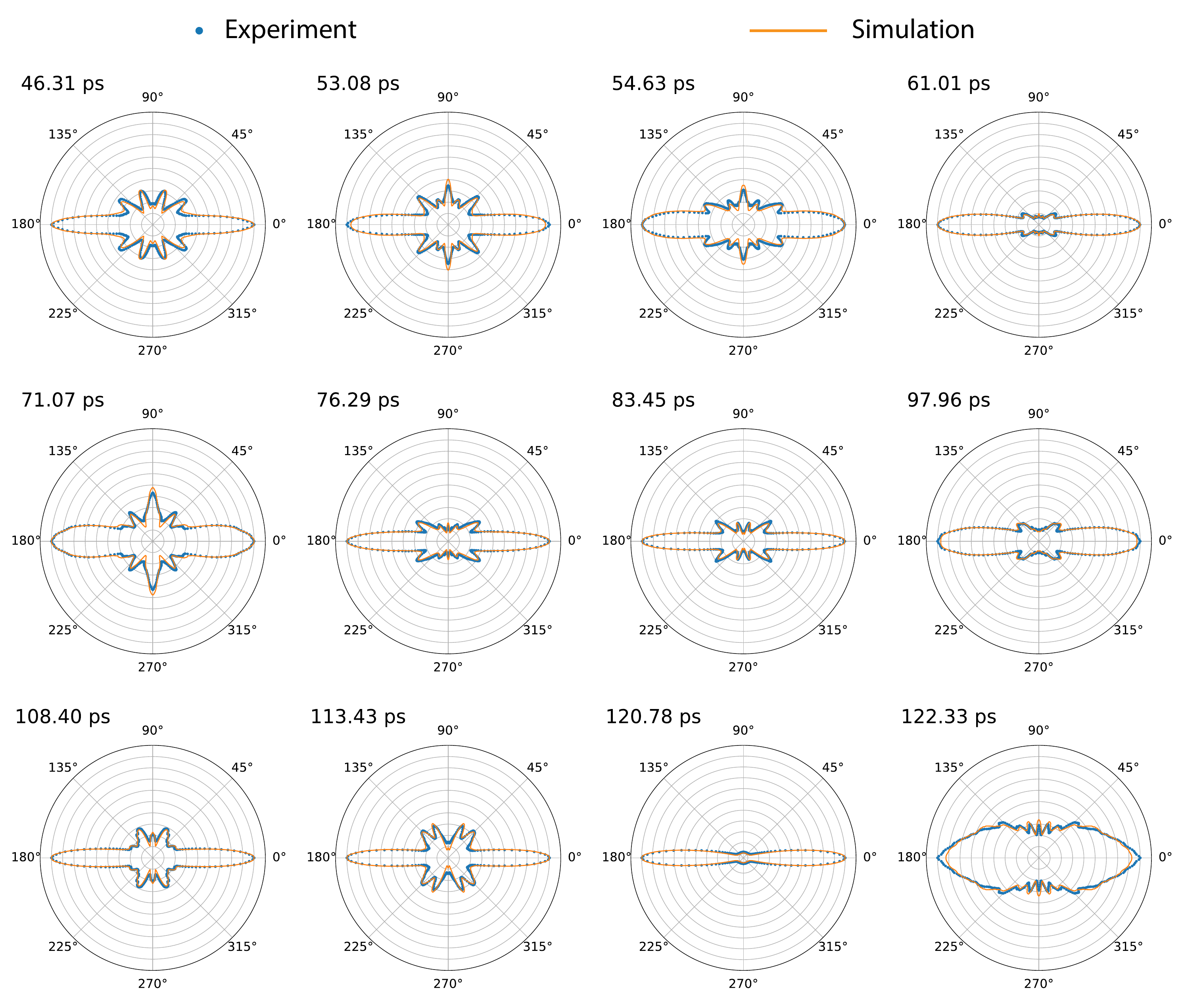}
   \caption{Comparison of experimental and simulated angular distributions for some selected delay
      times after the second alignment laser pulse has arrived.}
   \label{fig:comparison_ang_distributions}
\end{figure}
We show a comparison of angular distributions extracted from experiment and from the simulated, 2D
projected rotational densities for selected times, starting after the arrival of the second
alignment laser pulse in \autoref{fig:comparison_ang_distributions}. The angular distributions
display rich features, the simplest one being the alignment revival at a delay time of $120.78$~ps.
For better visibility the angular distributions have all been scaled up individually to maximise
visibility, except for the alignment revival at $120.78$~ps.

\section{Angular resolution}
\begin{figure}
   \includegraphics[width=\linewidth]{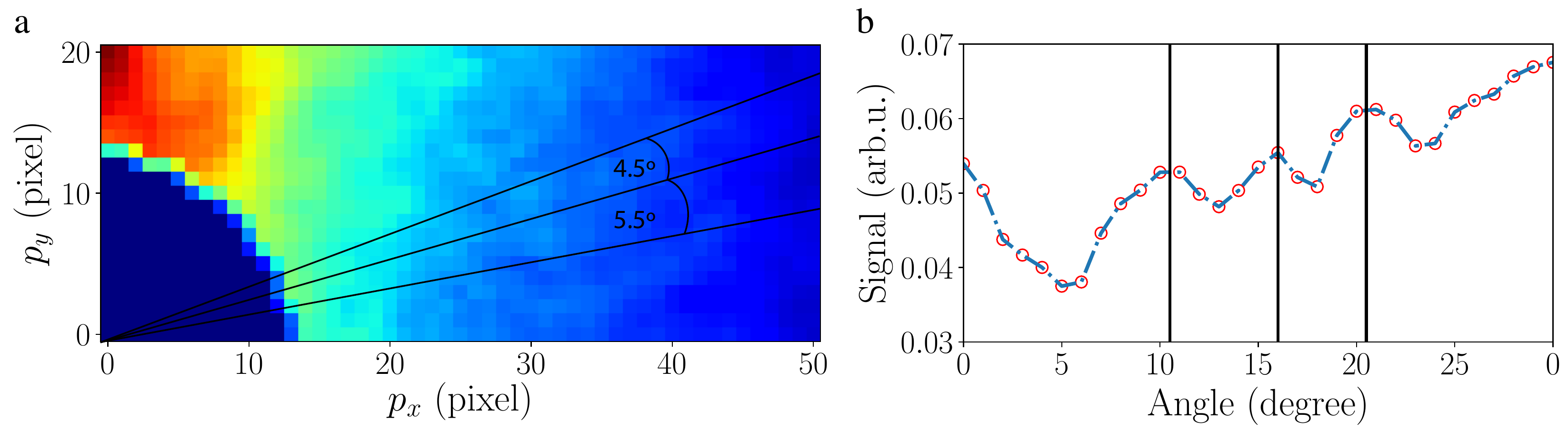}
   \caption{Determination of angular resolution. \textbf{a} Part of the recorded VMI image at a
      delay time of $98.5$~ps is shown with lines indicating the angles at which maxima in the
      angular distribution were observed. \textbf{b} Angular distribution for the same cutout of the
      VMI image, shown with the position of the maxima indicated by vertical lines. The smallest
      measured angle between maxima in the angular distribution is \degree{4.5}, close to the
      angular resolution of \degree{4}.}
   \label{fig:ang_resolution}
\end{figure}
The angular resolution was defined by the radius of the Coulomb channel in the VMI image, at which
the angular distribution was extracted, and the number of pixels needed to distinguish two
successive maxima or minima. The center of the radial Coulomb channel was at a radius of $46$ pixel,
which yielded an angle of \degree{1.26} per pixel, corresponding to a limit for the resolution to
separate two maxima or minima of \degree{4}. In \autoref[a]{fig:ang_resolution}, a O$^{+}$ ion
momentum distribution recorded at a delay time of $98.5$~ps is shown with lines indicating the
angles at which maxima in the angular distribution appear. In \autoref[b]{fig:ang_resolution} the
corresponding angular distribution is shown, where the maxima are clearly visible and
distinguishable.

\section{Highest observed degree of alignment}
\begin{figure}
   \includegraphics[width=\linewidth]{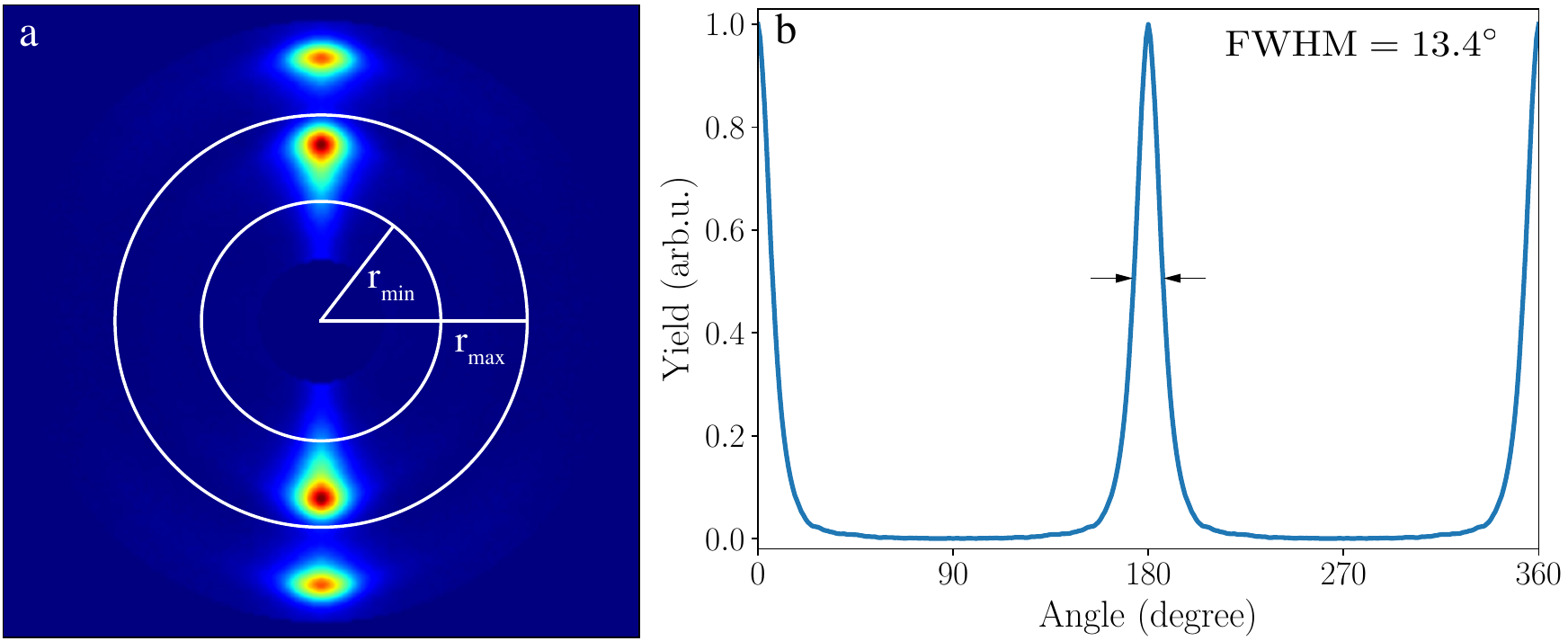}
   \caption{Observed highest degree of alignment of $\cost=0.955$. \textbf{a} O$^{+}$ ion momentum
      distribution recorded at the alignment revival at a delay time of $120.78$~ps. The integration
      for the calculation of \cost was carried out in the shell between $r_{\text{min}}=40$ and
      $r_{\text{max}}=64$ pixel. \textbf{b}~Corresponding angular distribution with a full opening
      angle of $\text{FWHM}=\degree{13.4}$}
   \label{fig:max_align}
\end{figure}
In \autoref[a]{fig:max_align} the O$^{+}$ ion momentum distribution of the strongest observed
field-free alignment is shown. The image was recorded at a delay time of $120.78$~ps, which is the
alignment revival, one rotational period after the arrival of the second alignment pulse. The degree
of alignment was $\cost=0.96$ as stated in the main text. The value was obtained through integration
in the shell between $r_{\text{min}}=40$ and $r_{\text{max}}=64$. In \autoref[b]{fig:max_align} the
corresponding angular distribution is shown, which yielded an opening angle of
$\text{FWHM}=\degree{13.4}$.

\subsection*{Acknowledgments}
This work has been supported by the Deutsche Forschungsgemeinschaft (DFG) through the priority
program ``Quantum Dynamics in Tailored Intense Fields'' (QUTIF, SPP1840, KU 1527/3, RO4577/4) and by
the Clusters of Excellence ``Center for Ultrafast Imaging'' (CUI, EXC~1074, ID~194651731) and
``Advanced Imaging of Matter'' (AIM, EXC~2056, ID~390715994) of the Deutsche Forschungsgemeinschaft
(DFG), and by the European Research Council under the European Union's Seventh Framework Programme
(FP7/2007-2013) through the Advanced Grant ``DropletControl'' (ERC-320459-Stapelfeldt) and the
Consolidator Grant ``COMOTION'' (ERC-614507-Küpper).

\section*{Author contributions}
The project was conceived and coordinated by AR and JK. The experiment was designed by KD, JK, AR;
set up by ETK, SR, KD, RRJ, and AR; and performed by ETK, SR, AT, ST, GG, PS, and AR. The data
analysis and numerical simulations were performed by ETK, the results from theory and experiment
were analyzed by ETK, TM, AT, ST, AR, and JK and discussed with HS and MJJV. The manuscript was
prepared by ETK, ST, and JK and discussed by all authors.

\section*{Competing Interests}
The authors declare that they have no competing financial interests.

\section*{Correspondence}
Correspondence and requests for materials should be addressed to Jochen Küpper
<jochen.kuepper@cfel.de> and Arnaud Rouz\'ee <rouzee@mbi-berlin.de>.

\bibliography{string,cmi}

\begin{thebibliography}{40}%
\makeatletter
\providecommand \@ifxundefined [1]{%
 \@ifx{#1\undefined}
}%
\providecommand \@ifnum [1]{%
 \ifnum #1\expandafter \@firstoftwo
 \else \expandafter \@secondoftwo
 \fi
}%
\providecommand \@ifx [1]{%
 \ifx #1\expandafter \@firstoftwo
 \else \expandafter \@secondoftwo
 \fi
}%
\providecommand \natexlab [1]{#1}%
\providecommand \enquote  [1]{``#1''}%
\providecommand \bibnamefont  [1]{#1}%
\providecommand \bibfnamefont [1]{#1}%
\providecommand \citenamefont [1]{#1}%
\providecommand \href@noop [0]{\@secondoftwo}%
\providecommand \href [0]{\begingroup \@sanitize@url \@href}%
\providecommand \@href[1]{\@@startlink{#1}\@@href}%
\providecommand \@@href[1]{\endgroup#1\@@endlink}%
\providecommand \@sanitize@url [0]{\catcode `\\12\catcode `\$12\catcode
  `\&12\catcode `\#12\catcode `\^12\catcode `\_12\catcode `\%12\relax}%
\providecommand \@@startlink[1]{}%
\providecommand \@@endlink[0]{}%
\providecommand \url  [0]{\begingroup\@sanitize@url \@url }%
\providecommand \@url [1]{\endgroup\@href {#1}{\urlprefix }}%
\providecommand \urlprefix  [0]{URL }%
\providecommand \Eprint [0]{\href }%
\providecommand \doibase [0]{http://dx.doi.org/}%
\providecommand \selectlanguage [0]{\@gobble}%
\providecommand \bibinfo  [0]{\@secondoftwo}%
\providecommand \bibfield  [0]{\@secondoftwo}%
\providecommand \translation [1]{[#1]}%
\providecommand \BibitemOpen [0]{}%
\providecommand \bibitemStop [0]{}%
\providecommand \bibitemNoStop [0]{.\EOS\space}%
\providecommand \EOS [0]{\spacefactor3000\relax}%
\providecommand \BibitemShut  [1]{\csname bibitem#1\endcsname}%
\let\auto@bib@innerbib\@empty
\bibitem [{\citenamefont {Zewail}(2000)}]{Zewail:JPCA104:5660}%
  \BibitemOpen
  \bibfield  {author} {\bibinfo {author} {\bibfnamefont {A.~H.}\ \bibnamefont
  {Zewail}},\ }\bibfield  {title} {\enquote {\bibinfo {title}
  {{Femtochemistry}: Atomic-scale dynamics of the chemical bond},}\ }\href
  {\doibase 10.1021/jp001460h} {\bibfield  {journal} {\bibinfo  {journal} {J.\
  Phys.\ Chem.\ A}\ }\textbf {\bibinfo {volume} {104}},\ \bibinfo {pages}
  {5660--5694} (\bibinfo {year} {2000})}\BibitemShut {NoStop}%
\bibitem [{\citenamefont {Ischenko}\ \emph {et~al.}(2017)\citenamefont
  {Ischenko}, \citenamefont {Weber},\ and\ \citenamefont
  {Miller}}]{Ischenko:CR117:11066}%
  \BibitemOpen
  \bibfield  {author} {\bibinfo {author} {\bibfnamefont {A.~A.}\ \bibnamefont
  {Ischenko}}, \bibinfo {author} {\bibfnamefont {P.~M.}\ \bibnamefont {Weber}},
  \ and\ \bibinfo {author} {\bibfnamefont {R.~J.~D.}\ \bibnamefont {Miller}},\
  }\bibfield  {title} {\enquote {\bibinfo {title} {Capturing chemistry in
  action with electrons: Realization of atomically resolved reaction
  dynamics},}\ }\href {\doibase 10.1021/acs.chemrev.6b00770} {\bibfield
  {journal} {\bibinfo  {journal} {Chem.\ Rev.}\ }\textbf {\bibinfo {volume}
  {117}},\ \bibinfo {pages} {11066--11124} (\bibinfo {year}
  {2017})}\BibitemShut {NoStop}%
\bibitem [{\citenamefont {Ayyer}\ \emph {et~al.}(2016)\citenamefont {Ayyer},
  \citenamefont {Yefanov}, \citenamefont {Oberth{\"u}r}, \citenamefont
  {Roy-Chowdhury}, \citenamefont {Galli}, \citenamefont {Mariani},
  \citenamefont {Basu}, \citenamefont {Coe}, \citenamefont {Conrad},
  \citenamefont {Fromme}, \citenamefont {Schaffer}, \citenamefont {D{\"o}rner},
  \citenamefont {James}, \citenamefont {Kupitz}, \citenamefont {Metz},
  \citenamefont {Nelson}, \citenamefont {Xavier}, \citenamefont {Beyerlein},
  \citenamefont {Schmidt}, \citenamefont {Sarrou}, \citenamefont {Spence},
  \citenamefont {Weierstall}, \citenamefont {White}, \citenamefont {Yang},
  \citenamefont {Zhao}, \citenamefont {Liang}, \citenamefont {Aquila},
  \citenamefont {Hunter}, \citenamefont {Robinson}, \citenamefont {Koglin},
  \citenamefont {Boutet}, \citenamefont {Fromme}, \citenamefont {Barty},\ and\
  \citenamefont {Chapman}}]{Ayyer:Nature530:202}%
  \BibitemOpen
  \bibfield  {author} {\bibinfo {author} {\bibfnamefont {K.}~\bibnamefont
  {Ayyer}}, \bibinfo {author} {\bibfnamefont {O.~M.}\ \bibnamefont {Yefanov}},
  \bibinfo {author} {\bibfnamefont {D.}~\bibnamefont {Oberth{\"u}r}}, \bibinfo
  {author} {\bibfnamefont {S.}~\bibnamefont {Roy-Chowdhury}}, \bibinfo {author}
  {\bibfnamefont {L.}~\bibnamefont {Galli}}, \bibinfo {author} {\bibfnamefont
  {V.}~\bibnamefont {Mariani}}, \bibinfo {author} {\bibfnamefont
  {S.}~\bibnamefont {Basu}}, \bibinfo {author} {\bibfnamefont {J.}~\bibnamefont
  {Coe}}, \bibinfo {author} {\bibfnamefont {C.~E.}\ \bibnamefont {Conrad}},
  \bibinfo {author} {\bibfnamefont {R.}~\bibnamefont {Fromme}}, \bibinfo
  {author} {\bibfnamefont {A.}~\bibnamefont {Schaffer}}, \bibinfo {author}
  {\bibfnamefont {K.}~\bibnamefont {D{\"o}rner}}, \bibinfo {author}
  {\bibfnamefont {D.}~\bibnamefont {James}}, \bibinfo {author} {\bibfnamefont
  {C.}~\bibnamefont {Kupitz}}, \bibinfo {author} {\bibfnamefont
  {M.}~\bibnamefont {Metz}}, \bibinfo {author} {\bibfnamefont {G.}~\bibnamefont
  {Nelson}}, \bibinfo {author} {\bibfnamefont {P.~L.}\ \bibnamefont {Xavier}},
  \bibinfo {author} {\bibfnamefont {K.~R.}\ \bibnamefont {Beyerlein}}, \bibinfo
  {author} {\bibfnamefont {M.}~\bibnamefont {Schmidt}}, \bibinfo {author}
  {\bibfnamefont {I.}~\bibnamefont {Sarrou}}, \bibinfo {author} {\bibfnamefont
  {J.~C.~H.}\ \bibnamefont {Spence}}, \bibinfo {author} {\bibfnamefont
  {U.}~\bibnamefont {Weierstall}}, \bibinfo {author} {\bibfnamefont {T.~A.}\
  \bibnamefont {White}}, \bibinfo {author} {\bibfnamefont {J.-H.}\ \bibnamefont
  {Yang}}, \bibinfo {author} {\bibfnamefont {Y.}~\bibnamefont {Zhao}}, \bibinfo
  {author} {\bibfnamefont {M.}~\bibnamefont {Liang}}, \bibinfo {author}
  {\bibfnamefont {A.}~\bibnamefont {Aquila}}, \bibinfo {author} {\bibfnamefont
  {M.~S.}\ \bibnamefont {Hunter}}, \bibinfo {author} {\bibfnamefont {J.~S.}\
  \bibnamefont {Robinson}}, \bibinfo {author} {\bibfnamefont {J.~E.}\
  \bibnamefont {Koglin}}, \bibinfo {author} {\bibfnamefont {S.}~\bibnamefont
  {Boutet}}, \bibinfo {author} {\bibfnamefont {P.}~\bibnamefont {Fromme}},
  \bibinfo {author} {\bibfnamefont {A.}~\bibnamefont {Barty}}, \ and\ \bibinfo
  {author} {\bibfnamefont {H.~N.}\ \bibnamefont {Chapman}},\ }\bibfield
  {title} {\enquote {\bibinfo {title} {Macromolecular diffractive imaging using
  imperfect crystals},}\ }\href {\doibase 10.1038/nature16949} {\bibfield
  {journal} {\bibinfo  {journal} {Nature}\ }\textbf {\bibinfo {volume} {530}},\
  \bibinfo {pages} {202--206} (\bibinfo {year} {2016})}\BibitemShut {NoStop}%
\bibitem [{\citenamefont {K{\"u}pper}\ \emph {et~al.}(2014)\citenamefont
  {K{\"u}pper}, \citenamefont {Stern}, \citenamefont {Holmegaard},
  \citenamefont {Filsinger}, \citenamefont {Rouz\'{e}e}, \citenamefont
  {Rudenko}, \citenamefont {Johnsson}, \citenamefont {Martin}, \citenamefont
  {Adolph}, \citenamefont {Aquila}, \citenamefont {Bajt}, \citenamefont
  {Barty}, \citenamefont {Bostedt}, \citenamefont {Bozek}, \citenamefont
  {Caleman}, \citenamefont {Coffee}, \citenamefont {Coppola}, \citenamefont
  {Delmas}, \citenamefont {Epp}, \citenamefont {Erk}, \citenamefont {Foucar},
  \citenamefont {Gorkhover}, \citenamefont {Gumprecht}, \citenamefont
  {Hartmann}, \citenamefont {Hartmann}, \citenamefont {Hauser}, \citenamefont
  {Holl}, \citenamefont {H{\"o}mke}, \citenamefont {Kimmel}, \citenamefont
  {Krasniqi}, \citenamefont {K{\"u}hnel}, \citenamefont {Maurer}, \citenamefont
  {Messerschmidt}, \citenamefont {Moshammer}, \citenamefont {Reich},
  \citenamefont {Rudek}, \citenamefont {Santra}, \citenamefont {Schlichting},
  \citenamefont {Schmidt}, \citenamefont {Schorb}, \citenamefont {Schulz},
  \citenamefont {Soltau}, \citenamefont {Spence}, \citenamefont {Starodub},
  \citenamefont {Str{\"u}der}, \citenamefont {Th{\o}gersen}, \citenamefont
  {Vrakking}, \citenamefont {Weidenspointner}, \citenamefont {White},
  \citenamefont {Wunderer}, \citenamefont {Meijer}, \citenamefont {Ullrich},
  \citenamefont {Stapelfeldt}, \citenamefont {Rolles},\ and\ \citenamefont
  {Chapman}}]{Kuepper:PRL112:083002}%
  \BibitemOpen
  \bibfield  {author} {\bibinfo {author} {\bibfnamefont {J.}~\bibnamefont
  {K{\"u}pper}}, \bibinfo {author} {\bibfnamefont {S.}~\bibnamefont {Stern}},
  \bibinfo {author} {\bibfnamefont {L.}~\bibnamefont {Holmegaard}}, \bibinfo
  {author} {\bibfnamefont {F.}~\bibnamefont {Filsinger}}, \bibinfo {author}
  {\bibfnamefont {A.}~\bibnamefont {Rouz\'{e}e}}, \bibinfo {author}
  {\bibfnamefont {A.}~\bibnamefont {Rudenko}}, \bibinfo {author} {\bibfnamefont
  {P.}~\bibnamefont {Johnsson}}, \bibinfo {author} {\bibfnamefont {A.~V.}\
  \bibnamefont {Martin}}, \bibinfo {author} {\bibfnamefont {M.}~\bibnamefont
  {Adolph}}, \bibinfo {author} {\bibfnamefont {A.}~\bibnamefont {Aquila}},
  \bibinfo {author} {\bibfnamefont {S.}~\bibnamefont {Bajt}}, \bibinfo {author}
  {\bibfnamefont {A.}~\bibnamefont {Barty}}, \bibinfo {author} {\bibfnamefont
  {C.}~\bibnamefont {Bostedt}}, \bibinfo {author} {\bibfnamefont
  {J.}~\bibnamefont {Bozek}}, \bibinfo {author} {\bibfnamefont
  {C.}~\bibnamefont {Caleman}}, \bibinfo {author} {\bibfnamefont
  {R.}~\bibnamefont {Coffee}}, \bibinfo {author} {\bibfnamefont
  {N.}~\bibnamefont {Coppola}}, \bibinfo {author} {\bibfnamefont
  {T.}~\bibnamefont {Delmas}}, \bibinfo {author} {\bibfnamefont
  {S.}~\bibnamefont {Epp}}, \bibinfo {author} {\bibfnamefont {B.}~\bibnamefont
  {Erk}}, \bibinfo {author} {\bibfnamefont {L.}~\bibnamefont {Foucar}},
  \bibinfo {author} {\bibfnamefont {T.}~\bibnamefont {Gorkhover}}, \bibinfo
  {author} {\bibfnamefont {L.}~\bibnamefont {Gumprecht}}, \bibinfo {author}
  {\bibfnamefont {A.}~\bibnamefont {Hartmann}}, \bibinfo {author}
  {\bibfnamefont {R.}~\bibnamefont {Hartmann}}, \bibinfo {author}
  {\bibfnamefont {G.}~\bibnamefont {Hauser}}, \bibinfo {author} {\bibfnamefont
  {P.}~\bibnamefont {Holl}}, \bibinfo {author} {\bibfnamefont {A.}~\bibnamefont
  {H{\"o}mke}}, \bibinfo {author} {\bibfnamefont {N.}~\bibnamefont {Kimmel}},
  \bibinfo {author} {\bibfnamefont {F.}~\bibnamefont {Krasniqi}}, \bibinfo
  {author} {\bibfnamefont {K.-U.}\ \bibnamefont {K{\"u}hnel}}, \bibinfo
  {author} {\bibfnamefont {J.}~\bibnamefont {Maurer}}, \bibinfo {author}
  {\bibfnamefont {M.}~\bibnamefont {Messerschmidt}}, \bibinfo {author}
  {\bibfnamefont {R.}~\bibnamefont {Moshammer}}, \bibinfo {author}
  {\bibfnamefont {C.}~\bibnamefont {Reich}}, \bibinfo {author} {\bibfnamefont
  {B.}~\bibnamefont {Rudek}}, \bibinfo {author} {\bibfnamefont
  {R.}~\bibnamefont {Santra}}, \bibinfo {author} {\bibfnamefont
  {I.}~\bibnamefont {Schlichting}}, \bibinfo {author} {\bibfnamefont
  {C.}~\bibnamefont {Schmidt}}, \bibinfo {author} {\bibfnamefont
  {S.}~\bibnamefont {Schorb}}, \bibinfo {author} {\bibfnamefont
  {J.}~\bibnamefont {Schulz}}, \bibinfo {author} {\bibfnamefont
  {H.}~\bibnamefont {Soltau}}, \bibinfo {author} {\bibfnamefont {J.~C.~H.}\
  \bibnamefont {Spence}}, \bibinfo {author} {\bibfnamefont {D.}~\bibnamefont
  {Starodub}}, \bibinfo {author} {\bibfnamefont {L.}~\bibnamefont
  {Str{\"u}der}}, \bibinfo {author} {\bibfnamefont {J.}~\bibnamefont
  {Th{\o}gersen}}, \bibinfo {author} {\bibfnamefont {M.~J.~J.}\ \bibnamefont
  {Vrakking}}, \bibinfo {author} {\bibfnamefont {G.}~\bibnamefont
  {Weidenspointner}}, \bibinfo {author} {\bibfnamefont {T.~A.}\ \bibnamefont
  {White}}, \bibinfo {author} {\bibfnamefont {C.}~\bibnamefont {Wunderer}},
  \bibinfo {author} {\bibfnamefont {G.}~\bibnamefont {Meijer}}, \bibinfo
  {author} {\bibfnamefont {J.}~\bibnamefont {Ullrich}}, \bibinfo {author}
  {\bibfnamefont {H.}~\bibnamefont {Stapelfeldt}}, \bibinfo {author}
  {\bibfnamefont {D.}~\bibnamefont {Rolles}}, \ and\ \bibinfo {author}
  {\bibfnamefont {H.~N.}\ \bibnamefont {Chapman}},\ }\bibfield  {title}
  {\enquote {\bibinfo {title} {X-ray diffraction from isolated and strongly
  aligned gas-phase molecules with a free-electron laser},}\ }\href {\doibase  10.1103/PhysRevLett.112.083002} {\bibfield  {journal} {\bibinfo  {journal}
  {Phys.\ Rev.\ Lett.}\ }\textbf {\bibinfo {volume} {112}},\ \bibinfo {pages}
  {083002} (\bibinfo {year} {2014})},\ \Eprint {http://arxiv.org/abs/1307.4577}
  {arXiv:1307.4577 [physics]} \BibitemShut {NoStop}%
\bibitem [{\citenamefont {Hensley}\ \emph {et~al.}(2012)\citenamefont
  {Hensley}, \citenamefont {Yang},\ and\ \citenamefont
  {Centurion}}]{Hensley:PRL109:133202}%
  \BibitemOpen
  \bibfield  {author} {\bibinfo {author} {\bibfnamefont {C.~J.}\ \bibnamefont
  {Hensley}}, \bibinfo {author} {\bibfnamefont {J.}~\bibnamefont {Yang}}, \
  and\ \bibinfo {author} {\bibfnamefont {M.}~\bibnamefont {Centurion}},\
  }\bibfield  {title} {\enquote {\bibinfo {title} {Imaging of isolated
  molecules with ultrafast electron pulses},}\ }\href {\doibase  10.1103/PhysRevLett.109.133202} {\bibfield  {journal} {\bibinfo  {journal}
  {Phys.\ Rev.\ Lett.}\ }\textbf {\bibinfo {volume} {109}},\ \bibinfo {pages}
  {133202} (\bibinfo {year} {2012})}\BibitemShut {NoStop}%
\bibitem [{\citenamefont {Pande}\ \emph {et~al.}(2016)\citenamefont {Pande},
  \citenamefont {Hutchison}, \citenamefont {Groenhof}, \citenamefont {Aquila},
  \citenamefont {Robinson}, \citenamefont {Tenboer}, \citenamefont {Basu},
  \citenamefont {Boutet}, \citenamefont {DePonte}, \citenamefont {Liang},
  \citenamefont {White}, \citenamefont {Zatsepin}, \citenamefont {Yefanov},
  \citenamefont {Morozov}, \citenamefont {Oberthuer}, \citenamefont {Gati},
  \citenamefont {Subramanian}, \citenamefont {James}, \citenamefont {Zhao},
  \citenamefont {Koralek}, \citenamefont {Brayshaw}, \citenamefont {Kupitz},
  \citenamefont {Conrad}, \citenamefont {Roy-Chowdhury}, \citenamefont {Coe},
  \citenamefont {Metz}, \citenamefont {Xavier}, \citenamefont {Grant},
  \citenamefont {Koglin}, \citenamefont {Ketawala}, \citenamefont {Fromme},
  \citenamefont {{\v S}rajer}, \citenamefont {Henning}, \citenamefont {Spence},
  \citenamefont {Ourmazd}, \citenamefont {Schwander}, \citenamefont
  {Weierstall}, \citenamefont {Frank}, \citenamefont {Fromme}, \citenamefont
  {Barty}, \citenamefont {Chapman}, \citenamefont {Moffat}, \citenamefont {van
  Thor},\ and\ \citenamefont {Schmidt}}]{Pande:Science352:725}%
  \BibitemOpen
  \bibfield  {author} {\bibinfo {author} {\bibfnamefont {K.}~\bibnamefont
  {Pande}}, \bibinfo {author} {\bibfnamefont {C.~D.~M.}\ \bibnamefont
  {Hutchison}}, \bibinfo {author} {\bibfnamefont {G.}~\bibnamefont {Groenhof}},
  \bibinfo {author} {\bibfnamefont {A.}~\bibnamefont {Aquila}}, \bibinfo
  {author} {\bibfnamefont {J.~S.}\ \bibnamefont {Robinson}}, \bibinfo {author}
  {\bibfnamefont {J.}~\bibnamefont {Tenboer}}, \bibinfo {author} {\bibfnamefont
  {S.}~\bibnamefont {Basu}}, \bibinfo {author} {\bibfnamefont {S.}~\bibnamefont
  {Boutet}}, \bibinfo {author} {\bibfnamefont {D.~P.}\ \bibnamefont {DePonte}},
  \bibinfo {author} {\bibfnamefont {M.}~\bibnamefont {Liang}}, \bibinfo
  {author} {\bibfnamefont {T.~A.}\ \bibnamefont {White}}, \bibinfo {author}
  {\bibfnamefont {N.~A.}\ \bibnamefont {Zatsepin}}, \bibinfo {author}
  {\bibfnamefont {O.}~\bibnamefont {Yefanov}}, \bibinfo {author} {\bibfnamefont
  {D.}~\bibnamefont {Morozov}}, \bibinfo {author} {\bibfnamefont
  {D.}~\bibnamefont {Oberthuer}}, \bibinfo {author} {\bibfnamefont
  {C.}~\bibnamefont {Gati}}, \bibinfo {author} {\bibfnamefont {G.}~\bibnamefont
  {Subramanian}}, \bibinfo {author} {\bibfnamefont {D.}~\bibnamefont {James}},
  \bibinfo {author} {\bibfnamefont {Y.}~\bibnamefont {Zhao}}, \bibinfo {author}
  {\bibfnamefont {J.}~\bibnamefont {Koralek}}, \bibinfo {author} {\bibfnamefont
  {J.}~\bibnamefont {Brayshaw}}, \bibinfo {author} {\bibfnamefont
  {C.}~\bibnamefont {Kupitz}}, \bibinfo {author} {\bibfnamefont
  {C.}~\bibnamefont {Conrad}}, \bibinfo {author} {\bibfnamefont
  {S.}~\bibnamefont {Roy-Chowdhury}}, \bibinfo {author} {\bibfnamefont {J.~D.}\
  \bibnamefont {Coe}}, \bibinfo {author} {\bibfnamefont {M.}~\bibnamefont
  {Metz}}, \bibinfo {author} {\bibfnamefont {P.~L.}\ \bibnamefont {Xavier}},
  \bibinfo {author} {\bibfnamefont {T.~D.}\ \bibnamefont {Grant}}, \bibinfo
  {author} {\bibfnamefont {J.~E.}\ \bibnamefont {Koglin}}, \bibinfo {author}
  {\bibfnamefont {G.}~\bibnamefont {Ketawala}}, \bibinfo {author}
  {\bibfnamefont {R.}~\bibnamefont {Fromme}}, \bibinfo {author} {\bibfnamefont
  {V.}~\bibnamefont {{\v S}rajer}}, \bibinfo {author} {\bibfnamefont
  {R.}~\bibnamefont {Henning}}, \bibinfo {author} {\bibfnamefont {J.~C.~H.}\
  \bibnamefont {Spence}}, \bibinfo {author} {\bibfnamefont {A.}~\bibnamefont
  {Ourmazd}}, \bibinfo {author} {\bibfnamefont {P.}~\bibnamefont {Schwander}},
  \bibinfo {author} {\bibfnamefont {U.}~\bibnamefont {Weierstall}}, \bibinfo
  {author} {\bibfnamefont {M.}~\bibnamefont {Frank}}, \bibinfo {author}
  {\bibfnamefont {P.}~\bibnamefont {Fromme}}, \bibinfo {author} {\bibfnamefont
  {A.}~\bibnamefont {Barty}}, \bibinfo {author} {\bibfnamefont {H.~N.}\
  \bibnamefont {Chapman}}, \bibinfo {author} {\bibfnamefont {K.}~\bibnamefont
  {Moffat}}, \bibinfo {author} {\bibfnamefont {J.~J.}\ \bibnamefont {van
  Thor}}, \ and\ \bibinfo {author} {\bibfnamefont {M.}~\bibnamefont
  {Schmidt}},\ }\bibfield  {title} {\enquote {\bibinfo {title} {Femtosecond
  structural dynamics drives the trans/cis isomerization in photoactive yellow
  protein},}\ }\href {\doibase 10.1126/science.aad5081} {\bibfield  {journal}
  {\bibinfo  {journal} {Science}\ }\textbf {\bibinfo {volume} {352}},\ \bibinfo
  {pages} {725--729} (\bibinfo {year} {2016})}\BibitemShut {NoStop}%
\bibitem [{\citenamefont {Yang}\ \emph
  {et~al.}(2016{\natexlab{a}})\citenamefont {Yang}, \citenamefont {Guehr},
  \citenamefont {Shen}, \citenamefont {Li}, \citenamefont {Vecchione},
  \citenamefont {Coffee}, \citenamefont {Corbett}, \citenamefont {Fry},
  \citenamefont {Hartmann}, \citenamefont {Hast}, \citenamefont {Hegazy},
  \citenamefont {Jobe}, \citenamefont {Makasyuk}, \citenamefont {Robinson},
  \citenamefont {Robinson}, \citenamefont {Vetter}, \citenamefont {Weathersby},
  \citenamefont {Yoneda}, \citenamefont {Wang},\ and\ \citenamefont
  {Centurion}}]{Yang:PRL117:153002}%
  \BibitemOpen
  \bibfield  {author} {\bibinfo {author} {\bibfnamefont {J.}~\bibnamefont
  {Yang}}, \bibinfo {author} {\bibfnamefont {M.}~\bibnamefont {Guehr}},
  \bibinfo {author} {\bibfnamefont {X.}~\bibnamefont {Shen}}, \bibinfo {author}
  {\bibfnamefont {R.}~\bibnamefont {Li}}, \bibinfo {author} {\bibfnamefont
  {T.}~\bibnamefont {Vecchione}}, \bibinfo {author} {\bibfnamefont
  {R.}~\bibnamefont {Coffee}}, \bibinfo {author} {\bibfnamefont
  {J.}~\bibnamefont {Corbett}}, \bibinfo {author} {\bibfnamefont
  {A.}~\bibnamefont {Fry}}, \bibinfo {author} {\bibfnamefont {N.}~\bibnamefont
  {Hartmann}}, \bibinfo {author} {\bibfnamefont {C.}~\bibnamefont {Hast}},
  \bibinfo {author} {\bibfnamefont {K.}~\bibnamefont {Hegazy}}, \bibinfo
  {author} {\bibfnamefont {K.}~\bibnamefont {Jobe}}, \bibinfo {author}
  {\bibfnamefont {I.}~\bibnamefont {Makasyuk}}, \bibinfo {author}
  {\bibfnamefont {J.}~\bibnamefont {Robinson}}, \bibinfo {author}
  {\bibfnamefont {M.~S.}\ \bibnamefont {Robinson}}, \bibinfo {author}
  {\bibfnamefont {S.}~\bibnamefont {Vetter}}, \bibinfo {author} {\bibfnamefont
  {S.}~\bibnamefont {Weathersby}}, \bibinfo {author} {\bibfnamefont
  {C.}~\bibnamefont {Yoneda}}, \bibinfo {author} {\bibfnamefont
  {X.}~\bibnamefont {Wang}}, \ and\ \bibinfo {author} {\bibfnamefont
  {M.}~\bibnamefont {Centurion}},\ }\bibfield  {title} {\enquote {\bibinfo
  {title} {Diffractive imaging of coherent nuclear motion in isolated
  molecules},}\ }\href {\doibase 10.1103/PhysRevLett.117.153002} {\bibfield
  {journal} {\bibinfo  {journal} {Phys.\ Rev.\ Lett.}\ }\textbf {\bibinfo
  {volume} {117}},\ \bibinfo {pages} {153002} (\bibinfo {year}
  {2016}{\natexlab{a}})}\BibitemShut {NoStop}%
\bibitem [{\citenamefont {Felker}\ \emph {et~al.}(1986)\citenamefont {Felker},
  \citenamefont {Baskin},\ and\ \citenamefont {Zewail}}]{Felker:JPC90:724}%
  \BibitemOpen
  \bibfield  {author} {\bibinfo {author} {\bibfnamefont {P.~M.}\ \bibnamefont
  {Felker}}, \bibinfo {author} {\bibfnamefont {J.~S.}\ \bibnamefont {Baskin}},
  \ and\ \bibinfo {author} {\bibfnamefont {A.~H.}\ \bibnamefont {Zewail}},\
  }\bibfield  {title} {\enquote {\bibinfo {title} {Rephasing of collisionless
  molecular rotational coherence in large molecules},}\ }\href {\doibase  10.1021/j100277a006} {\bibfield  {journal} {\bibinfo  {journal} {J.\ Phys.\
  Chem.}\ }\textbf {\bibinfo {volume} {90}},\ \bibinfo {pages} {724--728}
  (\bibinfo {year} {1986})}\BibitemShut {NoStop}%
\bibitem [{\citenamefont {Rosca-Pruna}\ and\ \citenamefont
  {Vrakking}(2001)}]{RoscaPruna:PRL87:153902}%
  \BibitemOpen
  \bibfield  {author} {\bibinfo {author} {\bibfnamefont {F.}~\bibnamefont
  {Rosca-Pruna}}\ and\ \bibinfo {author} {\bibfnamefont {M.~J.~J.}\
  \bibnamefont {Vrakking}},\ }\bibfield  {title} {\enquote {\bibinfo {title}
  {Experimental observation of revival structures in picosecond laser-induced
  alignment of {I}$_2$},}\ }\href {\doibase 10.1103/PhysRevLett.87.153902}
  {\bibfield  {journal} {\bibinfo  {journal} {Phys.\ Rev.\ Lett.}\ }\textbf
  {\bibinfo {volume} {87}},\ \bibinfo {pages} {153902} (\bibinfo {year}
  {2001})}\BibitemShut {NoStop}%
\bibitem [{\citenamefont {Stapelfeldt}\ and\ \citenamefont
  {Seideman}(2003)}]{Stapelfeldt:RMP75:543}%
  \BibitemOpen
  \bibfield  {author} {\bibinfo {author} {\bibfnamefont {H.}~\bibnamefont
  {Stapelfeldt}}\ and\ \bibinfo {author} {\bibfnamefont {T.}~\bibnamefont
  {Seideman}},\ }\bibfield  {title} {\enquote {\bibinfo {title} {Colloquium:
  Aligning molecules with strong laser pulses},}\ }\href {\doibase  10.1103/RevModPhys.75.543} {\bibfield  {journal} {\bibinfo  {journal} {Rev.\
  Mod.\ Phys.}\ }\textbf {\bibinfo {volume} {75}},\ \bibinfo {pages} {543--557}
  (\bibinfo {year} {2003})}\BibitemShut {NoStop}%
\bibitem [{\citenamefont {Mizuse}\ \emph {et~al.}(2015)\citenamefont {Mizuse},
  \citenamefont {Kitano}, \citenamefont {Hasegawa},\ and\ \citenamefont
  {Ohshima}}]{Mizuse:SciAdv1:e1400185}%
  \BibitemOpen
  \bibfield  {author} {\bibinfo {author} {\bibfnamefont {K.}~\bibnamefont
  {Mizuse}}, \bibinfo {author} {\bibfnamefont {K.}~\bibnamefont {Kitano}},
  \bibinfo {author} {\bibfnamefont {H.}~\bibnamefont {Hasegawa}}, \ and\
  \bibinfo {author} {\bibfnamefont {Y.}~\bibnamefont {Ohshima}},\ }\bibfield
  {title} {\enquote {\bibinfo {title} {Quantum unidirectional rotation directly
  imaged with molecules},}\ }\href {\doibase 10.1126/sciadv.1400185} {\bibfield
   {journal} {\bibinfo  {journal} {Science Advances}\ }\textbf {\bibinfo
  {volume} {1}},\ \bibinfo {pages} {e1400185} (\bibinfo {year}
  {2015})}\BibitemShut {NoStop}%
\bibitem [{\citenamefont {Dooley}\ \emph {et~al.}(2003)\citenamefont {Dooley},
  \citenamefont {Litvinyuk}, \citenamefont {Lee}, \citenamefont {Rayner},
  \citenamefont {Spanner}, \citenamefont {Villeneuve},\ and\ \citenamefont
  {Corkum}}]{Dooley:PRA68:023406}%
  \BibitemOpen
  \bibfield  {author} {\bibinfo {author} {\bibfnamefont {P.~W.}\ \bibnamefont
  {Dooley}}, \bibinfo {author} {\bibfnamefont {I.~V.}\ \bibnamefont
  {Litvinyuk}}, \bibinfo {author} {\bibfnamefont {K.~F.}\ \bibnamefont {Lee}},
  \bibinfo {author} {\bibfnamefont {D.~M.}\ \bibnamefont {Rayner}}, \bibinfo
  {author} {\bibfnamefont {M.}~\bibnamefont {Spanner}}, \bibinfo {author}
  {\bibfnamefont {D.~M.}\ \bibnamefont {Villeneuve}}, \ and\ \bibinfo {author}
  {\bibfnamefont {P.~B.}\ \bibnamefont {Corkum}},\ }\bibfield  {title}
  {\enquote {\bibinfo {title} {Direct imaging of rotational wave-packet
  dynamics of diatomic molecules},}\ }\href {\doibase  10.1103/PhysRevA.68.023406} {\bibfield  {journal} {\bibinfo  {journal}
  {Phys.\ Rev.\ A}\ }\textbf {\bibinfo {volume} {68}},\ \bibinfo {pages}
  {023406} (\bibinfo {year} {2003})}\BibitemShut {NoStop}%
\bibitem [{\citenamefont {Marceau}\ \emph {et~al.}(2017)\citenamefont
  {Marceau}, \citenamefont {Makhija}, \citenamefont {Platzer}, \citenamefont
  {Naumov}, \citenamefont {Corkum}, \citenamefont {Stolow}, \citenamefont
  {Villeneuve},\ and\ \citenamefont {Hockett}}]{Marceau:PRL119:083401}%
  \BibitemOpen
  \bibfield  {author} {\bibinfo {author} {\bibfnamefont {C.}~\bibnamefont
  {Marceau}}, \bibinfo {author} {\bibfnamefont {V.}~\bibnamefont {Makhija}},
  \bibinfo {author} {\bibfnamefont {D.}~\bibnamefont {Platzer}}, \bibinfo
  {author} {\bibfnamefont {A.~Y.}\ \bibnamefont {Naumov}}, \bibinfo {author}
  {\bibfnamefont {P.~B.}\ \bibnamefont {Corkum}}, \bibinfo {author}
  {\bibfnamefont {A.}~\bibnamefont {Stolow}}, \bibinfo {author} {\bibfnamefont
  {D.~M.}\ \bibnamefont {Villeneuve}}, \ and\ \bibinfo {author} {\bibfnamefont
  {P.}~\bibnamefont {Hockett}},\ }\bibfield  {title} {\enquote {\bibinfo
  {title} {Molecular frame reconstruction using time-domain photoionization
  interferometry},}\ }\href {\doibase 10.1103/PhysRevLett.119.083401}
  {\bibfield  {journal} {\bibinfo  {journal} {Phys.\ Rev.\ Lett.}\ }\textbf
  {\bibinfo {volume} {119}},\ \bibinfo {pages} {083401} (\bibinfo {year}
  {2017})}\BibitemShut {NoStop}%
\bibitem [{\citenamefont {Yang}\ \emph
  {et~al.}(2016{\natexlab{b}})\citenamefont {Yang}, \citenamefont {Guehr},
  \citenamefont {Vecchione}, \citenamefont {Robinson}, \citenamefont {Li},
  \citenamefont {Hartmann}, \citenamefont {Shen}, \citenamefont {Coffee},
  \citenamefont {Corbett}, \citenamefont {Fry}, \citenamefont {Gaffney},
  \citenamefont {Gorkhover}, \citenamefont {Hast}, \citenamefont {Jobe},
  \citenamefont {Makasyuk}, \citenamefont {Reid}, \citenamefont {Robinson},
  \citenamefont {Vetter}, \citenamefont {Wang}, \citenamefont {Weathersby},
  \citenamefont {Yoneda}, \citenamefont {Centurion},\ and\ \citenamefont
  {Wang}}]{Yang:NatComm7:11232}%
  \BibitemOpen
  \bibfield  {author} {\bibinfo {author} {\bibfnamefont {J.}~\bibnamefont
  {Yang}}, \bibinfo {author} {\bibfnamefont {M.}~\bibnamefont {Guehr}},
  \bibinfo {author} {\bibfnamefont {T.}~\bibnamefont {Vecchione}}, \bibinfo
  {author} {\bibfnamefont {M.~S.}\ \bibnamefont {Robinson}}, \bibinfo {author}
  {\bibfnamefont {R.}~\bibnamefont {Li}}, \bibinfo {author} {\bibfnamefont
  {N.}~\bibnamefont {Hartmann}}, \bibinfo {author} {\bibfnamefont
  {X.}~\bibnamefont {Shen}}, \bibinfo {author} {\bibfnamefont {R.}~\bibnamefont
  {Coffee}}, \bibinfo {author} {\bibfnamefont {J.}~\bibnamefont {Corbett}},
  \bibinfo {author} {\bibfnamefont {A.}~\bibnamefont {Fry}}, \bibinfo {author}
  {\bibfnamefont {K.}~\bibnamefont {Gaffney}}, \bibinfo {author} {\bibfnamefont
  {T.}~\bibnamefont {Gorkhover}}, \bibinfo {author} {\bibfnamefont
  {C.}~\bibnamefont {Hast}}, \bibinfo {author} {\bibfnamefont {K.}~\bibnamefont
  {Jobe}}, \bibinfo {author} {\bibfnamefont {I.}~\bibnamefont {Makasyuk}},
  \bibinfo {author} {\bibfnamefont {A.}~\bibnamefont {Reid}}, \bibinfo {author}
  {\bibfnamefont {J.}~\bibnamefont {Robinson}}, \bibinfo {author}
  {\bibfnamefont {S.}~\bibnamefont {Vetter}}, \bibinfo {author} {\bibfnamefont
  {F.}~\bibnamefont {Wang}}, \bibinfo {author} {\bibfnamefont {S.}~\bibnamefont
  {Weathersby}}, \bibinfo {author} {\bibfnamefont {C.}~\bibnamefont {Yoneda}},
  \bibinfo {author} {\bibfnamefont {M.}~\bibnamefont {Centurion}}, \ and\
  \bibinfo {author} {\bibfnamefont {X.}~\bibnamefont {Wang}},\ }\bibfield
  {title} {\enquote {\bibinfo {title} {Diffractive imaging of a rotational
  wavepacket in nitrogen molecules with femtosecond megaelectronvolt electron
  pulses},}\ }\href {\doibase 10.1038/ncomms11232} {\bibfield  {journal}
  {\bibinfo  {journal} {Nat. Commun.}\ }\textbf {\bibinfo {volume} {7}},\
  \bibinfo {pages} {11232} (\bibinfo {year} {2016}{\natexlab{b}})}\BibitemShut
  {NoStop}%
\bibitem [{\citenamefont {Ghafur}\ \emph {et~al.}(2009)\citenamefont {Ghafur},
  \citenamefont {Rouz\'ee}, \citenamefont {Gijsbertsen}, \citenamefont {Siu},
  \citenamefont {Stolte},\ and\ \citenamefont
  {Vrakking}}]{Ghafur:NatPhys5:289}%
  \BibitemOpen
  \bibfield  {author} {\bibinfo {author} {\bibfnamefont {O.}~\bibnamefont
  {Ghafur}}, \bibinfo {author} {\bibfnamefont {A.}~\bibnamefont {Rouz\'ee}},
  \bibinfo {author} {\bibfnamefont {A.}~\bibnamefont {Gijsbertsen}}, \bibinfo
  {author} {\bibfnamefont {W.~K.}\ \bibnamefont {Siu}}, \bibinfo {author}
  {\bibfnamefont {S.}~\bibnamefont {Stolte}}, \ and\ \bibinfo {author}
  {\bibfnamefont {M.~J.~J.}\ \bibnamefont {Vrakking}},\ }\bibfield  {title}
  {\enquote {\bibinfo {title} {Impulsive orientation and alignment of
  quantum-state-selected {NO} molecules},}\ }\href {\doibase 10.1038/nphys1225}
  {\bibfield  {journal} {\bibinfo  {journal} {Nat. Phys.}\ }\textbf {\bibinfo
  {volume} {5}},\ \bibinfo {pages} {289--293} (\bibinfo {year}
  {2009})}\BibitemShut {NoStop}%
\bibitem [{\citenamefont {Trippel}\ \emph {et~al.}(2015)\citenamefont
  {Trippel}, \citenamefont {Mullins}, \citenamefont {M{\"u}ller}, \citenamefont
  {Kienitz}, \citenamefont {Gonz{\'a}lez-F{\'e}rez},\ and\ \citenamefont
  {K{\"u}pper}}]{Trippel:PRL114:103003}%
  \BibitemOpen
  \bibfield  {author} {\bibinfo {author} {\bibfnamefont {S.}~\bibnamefont
  {Trippel}}, \bibinfo {author} {\bibfnamefont {T.}~\bibnamefont {Mullins}},
  \bibinfo {author} {\bibfnamefont {N.~L.~M.}\ \bibnamefont {M{\"u}ller}},
  \bibinfo {author} {\bibfnamefont {J.~S.}\ \bibnamefont {Kienitz}}, \bibinfo
  {author} {\bibfnamefont {R.}~\bibnamefont {Gonz{\'a}lez-F{\'e}rez}}, \ and\
  \bibinfo {author} {\bibfnamefont {J.}~\bibnamefont {K{\"u}pper}},\ }\bibfield
   {title} {\enquote {\bibinfo {title} {Two-state wave packet for strong
  field-free molecular orientation},}\ }\href {\doibase  10.1103/PhysRevLett.114.103003} {\bibfield  {journal} {\bibinfo  {journal}
  {Phys.\ Rev.\ Lett.}\ }\textbf {\bibinfo {volume} {114}},\ \bibinfo {pages}
  {103003} (\bibinfo {year} {2015})},\ \Eprint {http://arxiv.org/abs/1409.2836}
  {arXiv:1409.2836 [physics]} \BibitemShut {NoStop}%
\bibitem [{\citenamefont {Felker}(1992)}]{Felker:JPC96:7844}%
  \BibitemOpen
  \bibfield  {author} {\bibinfo {author} {\bibfnamefont {P.~M.}\ \bibnamefont
  {Felker}},\ }\bibfield  {title} {\enquote {\bibinfo {title} {Rotational
  coherence spectroscopy: studies of the geometries of large gas-phase species
  by picosecond time-domain methods},}\ }\href
  {http://pubs3.acs.org/acs/journals/doilookup?in_doi=10.1021/j100199a005}
  {\bibfield  {journal} {\bibinfo  {journal} {J.\ Phys.\ Chem.}\ }\textbf
  {\bibinfo {volume} {96}},\ \bibinfo {pages} {7844--7857} (\bibinfo {year}
  {1992})}\BibitemShut {NoStop}%
\bibitem [{\citenamefont {Riehn}(2002)}]{Riehn:CP283:297}%
  \BibitemOpen
  \bibfield  {author} {\bibinfo {author} {\bibfnamefont {C.}~\bibnamefont
  {Riehn}},\ }\bibfield  {title} {\enquote {\bibinfo {title} {High-resolution
  pump-probe rotational coherence spectroscopy - rotational constants and
  structure of ground and electronically excited states of large molecular
  systems},}\ }\href {\doibase 10.1016/S0301-0104(02)00572-4} {\bibfield
  {journal} {\bibinfo  {journal} {Chem.\ Phys.}\ }\textbf {\bibinfo {volume}
  {283}},\ \bibinfo {pages} {297--329} (\bibinfo {year} {2002})}\BibitemShut
  {NoStop}%
\bibitem [{\citenamefont {Lee}\ \emph {et~al.}(2004)\citenamefont {Lee},
  \citenamefont {Villeneuve}, \citenamefont {Corkum},\ and\ \citenamefont
  {Shapiro}}]{Lee:PRL93:233601}%
  \BibitemOpen
  \bibfield  {author} {\bibinfo {author} {\bibfnamefont {K.}~\bibnamefont
  {Lee}}, \bibinfo {author} {\bibfnamefont {D.}~\bibnamefont {Villeneuve}},
  \bibinfo {author} {\bibfnamefont {P.}~\bibnamefont {Corkum}}, \ and\ \bibinfo
  {author} {\bibfnamefont {E.}~\bibnamefont {Shapiro}},\ }\bibfield  {title}
  {\enquote {\bibinfo {title} {Phase control of rotational wave packets and
  quantum information},}\ }\href {\doibase 10.1103/PhysRevLett.93.233601}
  {\bibfield  {journal} {\bibinfo  {journal} {Phys.\ Rev.\ Lett.}\ }\textbf
  {\bibinfo {volume} {93}},\ \bibinfo {pages} {233601} (\bibinfo {year}
  {2004})}\BibitemShut {NoStop}%
\bibitem [{\citenamefont {Trippel}\ \emph {et~al.}(2014)\citenamefont
  {Trippel}, \citenamefont {Mullins}, \citenamefont {M{\"u}ller}, \citenamefont
  {Kienitz}, \citenamefont {Omiste}, \citenamefont {Stapelfeldt}, \citenamefont
  {Gonz{\'a}lez-F{\'e}rez},\ and\ \citenamefont
  {K{\"u}pper}}]{Trippel:PRA89:051401R}%
  \BibitemOpen
  \bibfield  {author} {\bibinfo {author} {\bibfnamefont {S.}~\bibnamefont
  {Trippel}}, \bibinfo {author} {\bibfnamefont {T.}~\bibnamefont {Mullins}},
  \bibinfo {author} {\bibfnamefont {N.~L.~M.}\ \bibnamefont {M{\"u}ller}},
  \bibinfo {author} {\bibfnamefont {J.~S.}\ \bibnamefont {Kienitz}}, \bibinfo
  {author} {\bibfnamefont {J.~J.}\ \bibnamefont {Omiste}}, \bibinfo {author}
  {\bibfnamefont {H.}~\bibnamefont {Stapelfeldt}}, \bibinfo {author}
  {\bibfnamefont {R.}~\bibnamefont {Gonz{\'a}lez-F{\'e}rez}}, \ and\ \bibinfo
  {author} {\bibfnamefont {J.}~\bibnamefont {K{\"u}pper}},\ }\bibfield  {title}
  {\enquote {\bibinfo {title} {Strongly driven quantum pendulum of the carbonyl
  sulfide molecule},}\ }\href {\doibase 10.1103/PhysRevA.89.051401} {\bibfield
  {journal} {\bibinfo  {journal} {Phys.\ Rev.\ A}\ }\textbf {\bibinfo {volume}
  {89}},\ \bibinfo {pages} {051401(R)} (\bibinfo {year} {2014})},\ \Eprint
  {http://arxiv.org/abs/1401.6897} {arXiv:1401.6897 [quant-ph]} \BibitemShut
  {NoStop}%
\bibitem [{\citenamefont {Mouritzen}\ and\ \citenamefont
  {Mølmer}(2006)}]{Mouritzen:JCP124:244311}%
  \BibitemOpen
  \bibfield  {author} {\bibinfo {author} {\bibfnamefont {A.~S.}\ \bibnamefont
  {Mouritzen}}\ and\ \bibinfo {author} {\bibfnamefont {K.}~\bibnamefont
  {Mølmer}},\ }\bibfield  {title} {\enquote {\bibinfo {title} {Quantum state
  tomography of molecular rotation},}\ }\href {\doibase 10.1063/1.2208351}
  {\bibfield  {journal} {\bibinfo  {journal} {J.\ Chem.\ Phys.}\ }\textbf
  {\bibinfo {volume} {124}},\ \bibinfo {pages} {244311} (\bibinfo {year}
  {2006})}\BibitemShut {NoStop}%
\bibitem [{\citenamefont {Hasegawa}\ and\ \citenamefont
  {Ohshima}(2008)}]{Hasegawa:PRL101:053002}%
  \BibitemOpen
  \bibfield  {author} {\bibinfo {author} {\bibfnamefont {H.}~\bibnamefont
  {Hasegawa}}\ and\ \bibinfo {author} {\bibfnamefont {Y.}~\bibnamefont
  {Ohshima}},\ }\bibfield  {title} {\enquote {\bibinfo {title} {Quantum state
  reconstruction of a rotational wave packet created by a nonresonant intense
  femtosecond laser field},}\ }\href {\doibase 10.1103/PhysRevLett.101.053002}
  {\bibfield  {journal} {\bibinfo  {journal} {Phys.\ Rev.\ Lett.}\ }\textbf
  {\bibinfo {volume} {101}},\ \bibinfo {pages} {053002} (\bibinfo {year}
  {2008})}\BibitemShut {NoStop}%
\bibitem [{\citenamefont {Berry}\ \emph {et~al.}(2001)\citenamefont {Berry},
  \citenamefont {Marzoli},\ and\ \citenamefont
  {Schleich}}]{Berry:PhysTod14:39}%
  \BibitemOpen
  \bibfield  {author} {\bibinfo {author} {\bibfnamefont {M.}~\bibnamefont
  {Berry}}, \bibinfo {author} {\bibfnamefont {I.}~\bibnamefont {Marzoli}}, \
  and\ \bibinfo {author} {\bibfnamefont {W.}~\bibnamefont {Schleich}},\
  }\bibfield  {title} {\enquote {\bibinfo {title} {Quantum carpets, carpets of
  light},}\ }\href {\doibase 10.1088/2058-7058/14/6/30} {\bibfield  {journal}
  {\bibinfo  {journal} {Phys.\ World}\ }\textbf {\bibinfo {volume} {14}},\
  \bibinfo {pages} {39--46} (\bibinfo {year} {2001})}\BibitemShut {NoStop}%
\bibitem [{\citenamefont {Leibscher}\ \emph {et~al.}(2003)\citenamefont
  {Leibscher}, \citenamefont {Averbukh},\ and\ \citenamefont
  {Rabitz}}]{Leibscher:PRL90:213001}%
  \BibitemOpen
  \bibfield  {author} {\bibinfo {author} {\bibfnamefont {M.}~\bibnamefont
  {Leibscher}}, \bibinfo {author} {\bibfnamefont {I.}~\bibnamefont {Averbukh}},
  \ and\ \bibinfo {author} {\bibfnamefont {H.}~\bibnamefont {Rabitz}},\
  }\bibfield  {title} {\enquote {\bibinfo {title} {Molecular alignment by
  trains of short laser pulses},}\ }\href {\doibase  10.1103/PhysRevLett.90.213001} {\bibfield  {journal} {\bibinfo  {journal}
  {Phys.\ Rev.\ Lett.}\ }\textbf {\bibinfo {volume} {90}},\ \bibinfo {pages}
  {213001} (\bibinfo {year} {2003})}\BibitemShut {NoStop}%
\bibitem [{\citenamefont {Gu{\'e}rin}\ \emph {et~al.}(2008)\citenamefont
  {Gu{\'e}rin}, \citenamefont {Rouz{\'e}e},\ and\ \citenamefont
  {Hertz}}]{Guerin:PRA77:041404}%
  \BibitemOpen
  \bibfield  {author} {\bibinfo {author} {\bibfnamefont {S.}~\bibnamefont
  {Gu{\'e}rin}}, \bibinfo {author} {\bibfnamefont {A.}~\bibnamefont
  {Rouz{\'e}e}}, \ and\ \bibinfo {author} {\bibfnamefont {E.}~\bibnamefont
  {Hertz}},\ }\bibfield  {title} {\enquote {\bibinfo {title} {Ultimate
  field-free molecular alignment by combined adiabatic-impulsive field
  design},}\ }\href {\doibase 10.1103/PhysRevA.77.041404} {\bibfield  {journal}
  {\bibinfo  {journal} {Phys.\ Rev.\ A}\ }\textbf {\bibinfo {volume} {77}},\
  \bibinfo {pages} {041404} (\bibinfo {year} {2008})}\BibitemShut {NoStop}%
\bibitem [{\citenamefont {Chang}\ \emph {et~al.}(2015)\citenamefont {Chang},
  \citenamefont {Horke}, \citenamefont {Trippel},\ and\ \citenamefont
  {K{\"u}pper}}]{Chang:IRPC34:557}%
  \BibitemOpen
  \bibfield  {author} {\bibinfo {author} {\bibfnamefont {Y.-P.}\ \bibnamefont
  {Chang}}, \bibinfo {author} {\bibfnamefont {D.~A.}\ \bibnamefont {Horke}},
  \bibinfo {author} {\bibfnamefont {S.}~\bibnamefont {Trippel}}, \ and\
  \bibinfo {author} {\bibfnamefont {J.}~\bibnamefont {K{\"u}pper}},\ }\bibfield
   {title} {\enquote {\bibinfo {title} {Spatially-controlled complex molecules
  and their applications},}\ }\href {\doibase 10.1080/0144235X.2015.1077838}
  {\bibfield  {journal} {\bibinfo  {journal} {Int.\ Rev.\ Phys.\ Chem.}\
  }\textbf {\bibinfo {volume} {34}},\ \bibinfo {pages} {557--590} (\bibinfo
  {year} {2015})},\ \Eprint {http://arxiv.org/abs/1505.05632} {arXiv:1505.05632
  [physics]} \BibitemShut {NoStop}%
\bibitem [{\citenamefont {Eppink}\ and\ \citenamefont
  {Parker}(1997)}]{Eppink:RSI68:3477}%
  \BibitemOpen
  \bibfield  {author} {\bibinfo {author} {\bibfnamefont {A.~T. J.~B.}\
  \bibnamefont {Eppink}}\ and\ \bibinfo {author} {\bibfnamefont {D.~H.}\
  \bibnamefont {Parker}},\ }\bibfield  {title} {\enquote {\bibinfo {title}
  {Velocity map imaging of ions and electrons using electrostatic lenses:
  Application in photoelectron and photofragment ion imaging of molecular
  oxygen},}\ }\href {\doibase 10.1063/1.1148310} {\bibfield  {journal}
  {\bibinfo  {journal} {Rev.\ Sci.\ Instrum.}\ }\textbf {\bibinfo {volume}
  {68}},\ \bibinfo {pages} {3477--3484} (\bibinfo {year} {1997})}\BibitemShut
  {NoStop}%
\bibitem [{\citenamefont {Pitzer}\ \emph {et~al.}(2013)\citenamefont {Pitzer},
  \citenamefont {Kunitski}, \citenamefont {Johnson}, \citenamefont {Jahnke},
  \citenamefont {Sann}, \citenamefont {Sturm}, \citenamefont {Schmidt},
  \citenamefont {Schmidt-Böcking}, \citenamefont {Dörner}, \citenamefont
  {Stohner}, \citenamefont {Kiedrowski}, \citenamefont {Reggelin},
  \citenamefont {Marquardt}, \citenamefont {Schießer}, \citenamefont
  {Berger},\ and\ \citenamefont {Schöffler}}]{Pitzer:Science341:1096}%
  \BibitemOpen
  \bibfield  {author} {\bibinfo {author} {\bibfnamefont {M.}~\bibnamefont
  {Pitzer}}, \bibinfo {author} {\bibfnamefont {M.}~\bibnamefont {Kunitski}},
  \bibinfo {author} {\bibfnamefont {A.~S.}\ \bibnamefont {Johnson}}, \bibinfo
  {author} {\bibfnamefont {T.}~\bibnamefont {Jahnke}}, \bibinfo {author}
  {\bibfnamefont {H.}~\bibnamefont {Sann}}, \bibinfo {author} {\bibfnamefont
  {F.}~\bibnamefont {Sturm}}, \bibinfo {author} {\bibfnamefont {L.~P.~H.}\
  \bibnamefont {Schmidt}}, \bibinfo {author} {\bibfnamefont {H.}~\bibnamefont
  {Schmidt-Böcking}}, \bibinfo {author} {\bibfnamefont {R.}~\bibnamefont
  {Dörner}}, \bibinfo {author} {\bibfnamefont {J.}~\bibnamefont {Stohner}},
  \bibinfo {author} {\bibfnamefont {J.}~\bibnamefont {Kiedrowski}}, \bibinfo
  {author} {\bibfnamefont {M.}~\bibnamefont {Reggelin}}, \bibinfo {author}
  {\bibfnamefont {S.}~\bibnamefont {Marquardt}}, \bibinfo {author}
  {\bibfnamefont {A.}~\bibnamefont {Schießer}}, \bibinfo {author}
  {\bibfnamefont {R.}~\bibnamefont {Berger}}, \ and\ \bibinfo {author}
  {\bibfnamefont {M.~S.}\ \bibnamefont {Schöffler}},\ }\bibfield  {title}
  {\enquote {\bibinfo {title} {Direct determination of absolute molecular
  stereochemistry in gas phase by coulomb explosion imaging},}\ }\href
  {\doibase 10.1126/science.1240362} {\bibfield  {journal} {\bibinfo  {journal}
  {Science}\ }\textbf {\bibinfo {volume} {341}},\ \bibinfo {pages} {1096--1100}
  (\bibinfo {year} {2013})}\BibitemShut {NoStop}%
\bibitem [{\citenamefont {Christensen}\ \emph {et~al.}(2014)\citenamefont
  {Christensen}, \citenamefont {Nielsen}, \citenamefont {Brandt}, \citenamefont
  {Madsen}, \citenamefont {Madsen}, \citenamefont {Slater}, \citenamefont
  {Lauer}, \citenamefont {Brouard}, \citenamefont {Johansson}, \citenamefont
  {Shepperson},\ and\ \citenamefont {Stapelfeldt}}]{Christensen:PRL113:073005}%
  \BibitemOpen
  \bibfield  {author} {\bibinfo {author} {\bibfnamefont {L.}~\bibnamefont
  {Christensen}}, \bibinfo {author} {\bibfnamefont {J.~H.}\ \bibnamefont
  {Nielsen}}, \bibinfo {author} {\bibfnamefont {C.~B.}\ \bibnamefont {Brandt}},
  \bibinfo {author} {\bibfnamefont {C.~B.}\ \bibnamefont {Madsen}}, \bibinfo
  {author} {\bibfnamefont {L.~B.}\ \bibnamefont {Madsen}}, \bibinfo {author}
  {\bibfnamefont {C.~S.}\ \bibnamefont {Slater}}, \bibinfo {author}
  {\bibfnamefont {A.}~\bibnamefont {Lauer}}, \bibinfo {author} {\bibfnamefont
  {M.}~\bibnamefont {Brouard}}, \bibinfo {author} {\bibfnamefont {M.~P.}\
  \bibnamefont {Johansson}}, \bibinfo {author} {\bibfnamefont {B.}~\bibnamefont
  {Shepperson}}, \ and\ \bibinfo {author} {\bibfnamefont {H.}~\bibnamefont
  {Stapelfeldt}},\ }\bibfield  {title} {\enquote {\bibinfo {title} {Dynamic
  stark control of torsional motion by a pair of laser pulses},}\ }\href
  {\doibase 10.1103/PhysRevLett.113.073005} {\bibfield  {journal} {\bibinfo
  {journal} {Phys.\ Rev.\ Lett.}\ }\textbf {\bibinfo {volume} {113}},\ \bibinfo
  {pages} {073005} (\bibinfo {year} {2014})}\BibitemShut {NoStop}%
\bibitem [{\citenamefont {Owens}\ \emph {et~al.}(2018)\citenamefont {Owens},
  \citenamefont {Yachmenev}, \citenamefont {Yurchenko},\ and\ \citenamefont
  {K\"{u}pper}}]{Owens:PRL121:193201}%
  \BibitemOpen
  \bibfield  {author} {\bibinfo {author} {\bibfnamefont {A.}~\bibnamefont
  {Owens}}, \bibinfo {author} {\bibfnamefont {A.}~\bibnamefont {Yachmenev}},
  \bibinfo {author} {\bibfnamefont {S.~N.}\ \bibnamefont {Yurchenko}}, \ and\
  \bibinfo {author} {\bibfnamefont {J.}~\bibnamefont {K\"{u}pper}},\ }\bibfield
   {title} {\enquote {\bibinfo {title} {{C}limbing the {R}otational {L}adder to
  {C}hirality},}\ }\href {\doibase 10.1103/physrevlett.121.193201} {\bibfield
  {journal} {\bibinfo  {journal} {Phys.\ Rev.\ Lett.}\ }\textbf {\bibinfo
  {volume} {121}},\ \bibinfo {pages} {193201} (\bibinfo {year} {2018})},\
  \Eprint {http://arxiv.org/abs/1802.07803} {arXiv:1802.07803 [physics]}
  \BibitemShut {NoStop}%
\bibitem [{\citenamefont {Kuipers}\ \emph {et~al.}(1988)\citenamefont
  {Kuipers}, \citenamefont {Tenner}, \citenamefont {Kleyn},\ and\ \citenamefont
  {Stolte}}]{Kuipers:Nature334:420}%
  \BibitemOpen
  \bibfield  {author} {\bibinfo {author} {\bibfnamefont {E.~W.}\ \bibnamefont
  {Kuipers}}, \bibinfo {author} {\bibfnamefont {M.~G.}\ \bibnamefont {Tenner}},
  \bibinfo {author} {\bibfnamefont {A.}~\bibnamefont {Kleyn}}, \ and\ \bibinfo
  {author} {\bibfnamefont {S.}~\bibnamefont {Stolte}},\ }\bibfield  {title}
  {\enquote {\bibinfo {title} {Observation of steric effects in gas-surface
  scattering},}\ }\href {\doibase doi:10.1038/334420a0} {\bibfield  {journal}
  {\bibinfo  {journal} {Nature}\ }\textbf {\bibinfo {volume} {334}},\ \bibinfo
  {pages} {420--422} (\bibinfo {year} {1988})}\BibitemShut {NoStop}%
\bibitem [{\citenamefont {Rakitzis}\ \emph {et~al.}(2004)\citenamefont
  {Rakitzis}, \citenamefont {van~den Brom},\ and\ \citenamefont
  {Janssen}}]{Rakitzis:Science303:1852}%
  \BibitemOpen
  \bibfield  {author} {\bibinfo {author} {\bibfnamefont {T.~P.}\ \bibnamefont
  {Rakitzis}}, \bibinfo {author} {\bibfnamefont {A.~J.}\ \bibnamefont {van~den
  Brom}}, \ and\ \bibinfo {author} {\bibfnamefont {M.~H.~M.}\ \bibnamefont
  {Janssen}},\ }\bibfield  {title} {\enquote {\bibinfo {title} {Directional
  dynamics in the photodissociation of oriented molecules},}\ }\href {\doibase  10.1126/science.1094186} {\bibfield  {journal} {\bibinfo  {journal}
  {Science}\ }\textbf {\bibinfo {volume} {303}},\ \bibinfo {pages} {1852--1854}
  (\bibinfo {year} {2004})}\BibitemShut {NoStop}%
\bibitem [{\citenamefont {Itatani}\ \emph {et~al.}(2004)\citenamefont
  {Itatani}, \citenamefont {Levesque}, \citenamefont {Zeidler}, \citenamefont
  {Niikura}, \citenamefont {P\'{e}pin}, \citenamefont {Kieffer}, \citenamefont
  {Corkum},\ and\ \citenamefont {Villeneuve}}]{Itatani:Nature432:867}%
  \BibitemOpen
  \bibfield  {author} {\bibinfo {author} {\bibfnamefont {J.}~\bibnamefont
  {Itatani}}, \bibinfo {author} {\bibfnamefont {J.}~\bibnamefont {Levesque}},
  \bibinfo {author} {\bibfnamefont {D.}~\bibnamefont {Zeidler}}, \bibinfo
  {author} {\bibfnamefont {H.}~\bibnamefont {Niikura}}, \bibinfo {author}
  {\bibfnamefont {H.}~\bibnamefont {P\'{e}pin}}, \bibinfo {author}
  {\bibfnamefont {J.~C.}\ \bibnamefont {Kieffer}}, \bibinfo {author}
  {\bibfnamefont {P.~B.}\ \bibnamefont {Corkum}}, \ and\ \bibinfo {author}
  {\bibfnamefont {D.~M.}\ \bibnamefont {Villeneuve}},\ }\bibfield  {title}
  {\enquote {\bibinfo {title} {Tomographic imaging of molecular orbitals},}\
  }\href {\doibase 10.1038/nature03183} {\bibfield  {journal} {\bibinfo
  {journal} {Nature}\ }\textbf {\bibinfo {volume} {432}},\ \bibinfo {pages}
  {867--871} (\bibinfo {year} {2004})}\BibitemShut {NoStop}%
\bibitem [{\citenamefont {Holmegaard}\ \emph {et~al.}(2010)\citenamefont
  {Holmegaard}, \citenamefont {Hansen}, \citenamefont {Kalh{\o}j},
  \citenamefont {Kragh}, \citenamefont {Stapelfeldt}, \citenamefont
  {Filsinger}, \citenamefont {K{\"u}pper}, \citenamefont {Meijer},
  \citenamefont {Dimitrovski}, \citenamefont {Abu-samha}, \citenamefont
  {Martiny},\ and\ \citenamefont {Madsen}}]{Holmegaard:NatPhys6:428}%
  \BibitemOpen
  \bibfield  {author} {\bibinfo {author} {\bibfnamefont {L.}~\bibnamefont
  {Holmegaard}}, \bibinfo {author} {\bibfnamefont {J.~L.}\ \bibnamefont
  {Hansen}}, \bibinfo {author} {\bibfnamefont {L.}~\bibnamefont {Kalh{\o}j}},
  \bibinfo {author} {\bibfnamefont {S.~L.}\ \bibnamefont {Kragh}}, \bibinfo
  {author} {\bibfnamefont {H.}~\bibnamefont {Stapelfeldt}}, \bibinfo {author}
  {\bibfnamefont {F.}~\bibnamefont {Filsinger}}, \bibinfo {author}
  {\bibfnamefont {J.}~\bibnamefont {K{\"u}pper}}, \bibinfo {author}
  {\bibfnamefont {G.}~\bibnamefont {Meijer}}, \bibinfo {author} {\bibfnamefont
  {D.}~\bibnamefont {Dimitrovski}}, \bibinfo {author} {\bibfnamefont
  {M.}~\bibnamefont {Abu-samha}}, \bibinfo {author} {\bibfnamefont {C.~P.~J.}\
  \bibnamefont {Martiny}}, \ and\ \bibinfo {author} {\bibfnamefont {L.~B.}\
  \bibnamefont {Madsen}},\ }\bibfield  {title} {\enquote {\bibinfo {title}
  {Photoelectron angular distributions from strong-field ionization of oriented
  molecules},}\ }\href {\doibase 10.1038/NPHYS1666} {\bibfield  {journal}
  {\bibinfo  {journal} {Nat. Phys.}\ }\textbf {\bibinfo {volume} {6}},\
  \bibinfo {pages} {428} (\bibinfo {year} {2010})},\ \Eprint
  {http://arxiv.org/abs/1003.4634} {arXiv:1003.4634 [physics]} \BibitemShut
  {NoStop}%
\bibitem [{\citenamefont {Filsinger}\ \emph {et~al.}(2011)\citenamefont
  {Filsinger}, \citenamefont {Meijer}, \citenamefont {Stapelfeldt},
  \citenamefont {Chapman},\ and\ \citenamefont
  {K{\"u}pper}}]{Filsinger:PCCP13:2076}%
  \BibitemOpen
  \bibfield  {author} {\bibinfo {author} {\bibfnamefont {F.}~\bibnamefont
  {Filsinger}}, \bibinfo {author} {\bibfnamefont {G.}~\bibnamefont {Meijer}},
  \bibinfo {author} {\bibfnamefont {H.}~\bibnamefont {Stapelfeldt}}, \bibinfo
  {author} {\bibfnamefont {H.}~\bibnamefont {Chapman}}, \ and\ \bibinfo
  {author} {\bibfnamefont {J.}~\bibnamefont {K{\"u}pper}},\ }\bibfield  {title}
  {\enquote {\bibinfo {title} {State- and conformer-selected beams of aligned
  and oriented molecules for ultrafast diffraction studies},}\ }\href {\doibase  10.1039/C0CP01585G} {\bibfield  {journal} {\bibinfo  {journal} {Phys.\ Chem.\
  Chem.\ Phys.}\ }\textbf {\bibinfo {volume} {13}},\ \bibinfo {pages}
  {2076--2087} (\bibinfo {year} {2011})},\ \Eprint
  {http://arxiv.org/abs/1009.0871} {arXiv:1009.0871 [physics]} \BibitemShut
  {NoStop}%
\bibitem [{\citenamefont {Weber}\ \emph {et~al.}(2013)\citenamefont {Weber},
  \citenamefont {Oppermann},\ and\ \citenamefont
  {Marangos}}]{Weber:PRL111:263601}%
  \BibitemOpen
  \bibfield  {author} {\bibinfo {author} {\bibfnamefont {S.~J.}\ \bibnamefont
  {Weber}}, \bibinfo {author} {\bibfnamefont {M.}~\bibnamefont {Oppermann}}, \
  and\ \bibinfo {author} {\bibfnamefont {J.~P.}\ \bibnamefont {Marangos}},\
  }\bibfield  {title} {\enquote {\bibinfo {title} {Role of rotational wave
  packets in strong field experiments},}\ }\href {\doibase  10.1103/PhysRevLett.111.263601} {\bibfield  {journal} {\bibinfo  {journal}
  {Phys.\ Rev.\ Lett.}\ }\textbf {\bibinfo {volume} {111}},\ \bibinfo {pages}
  {263601} (\bibinfo {year} {2013})}\BibitemShut {NoStop}%
\bibitem [{\citenamefont {Pullen}\ \emph {et~al.}(2015)\citenamefont {Pullen},
  \citenamefont {Wolter}, \citenamefont {Le}, \citenamefont {Baudisch},
  \citenamefont {Hemmer}, \citenamefont {Senftleben}, \citenamefont {Schroter},
  \citenamefont {Ullrich}, \citenamefont {Moshammer}, \citenamefont {Lin},\
  and\ \citenamefont {Biegert}}]{Pullen:NatComm6:7262}%
  \BibitemOpen
  \bibfield  {author} {\bibinfo {author} {\bibfnamefont {M.~G.}\ \bibnamefont
  {Pullen}}, \bibinfo {author} {\bibfnamefont {B.}~\bibnamefont {Wolter}},
  \bibinfo {author} {\bibfnamefont {A.-T.}\ \bibnamefont {Le}}, \bibinfo
  {author} {\bibfnamefont {M.}~\bibnamefont {Baudisch}}, \bibinfo {author}
  {\bibfnamefont {M.}~\bibnamefont {Hemmer}}, \bibinfo {author} {\bibfnamefont
  {A.}~\bibnamefont {Senftleben}}, \bibinfo {author} {\bibfnamefont {C.~D.}\
  \bibnamefont {Schroter}}, \bibinfo {author} {\bibfnamefont {J.}~\bibnamefont
  {Ullrich}}, \bibinfo {author} {\bibfnamefont {R.}~\bibnamefont {Moshammer}},
  \bibinfo {author} {\bibfnamefont {C.~D.}\ \bibnamefont {Lin}}, \ and\
  \bibinfo {author} {\bibfnamefont {J.}~\bibnamefont {Biegert}},\ }\bibfield
  {title} {\enquote {\bibinfo {title} {Imaging an aligned polyatomic molecule
  with laser-induced electron diffraction},}\ }\href {\doibase  10.1038/ncomms8262} {\bibfield  {journal} {\bibinfo  {journal} {Nat.
  Commun.}\ }\textbf {\bibinfo {volume} {6}},\ \bibinfo {pages} {7262}
  (\bibinfo {year} {2015})}\BibitemShut {NoStop}%
\bibitem [{\citenamefont {Barty}\ \emph {et~al.}(2013)\citenamefont {Barty},
  \citenamefont {K{\"u}pper},\ and\ \citenamefont
  {Chapman}}]{Barty:ARPC64:415}%
  \BibitemOpen
  \bibfield  {author} {\bibinfo {author} {\bibfnamefont {A.}~\bibnamefont
  {Barty}}, \bibinfo {author} {\bibfnamefont {J.}~\bibnamefont {K{\"u}pper}}, \
  and\ \bibinfo {author} {\bibfnamefont {H.~N.}\ \bibnamefont {Chapman}},\
  }\bibfield  {title} {\enquote {\bibinfo {title} {Molecular imaging using
  x-ray free-electron lasers},}\ }\href {\doibase  10.1146/annurev-physchem-032511-143708} {\bibfield  {journal} {\bibinfo
  {journal} {Annu.\ Rev.\ Phys.\ Chem.}\ }\textbf {\bibinfo {volume} {64}},\
  \bibinfo {pages} {415--435} (\bibinfo {year} {2013})}\BibitemShut {NoStop}%
\bibitem [{\citenamefont {Hillenkamp}\ \emph {et~al.}(2003)\citenamefont
  {Hillenkamp}, \citenamefont {Keinan},\ and\ \citenamefont
  {Even}}]{Hillenkamp:JCP118:8699}%
  \BibitemOpen
  \bibfield  {author} {\bibinfo {author} {\bibfnamefont {M.}~\bibnamefont
  {Hillenkamp}}, \bibinfo {author} {\bibfnamefont {S.}~\bibnamefont {Keinan}},
  \ and\ \bibinfo {author} {\bibfnamefont {U.}~\bibnamefont {Even}},\
  }\bibfield  {title} {\enquote {\bibinfo {title} {Condensation limited cooling
  in supersonic expansions},}\ }\href {\doibase 10.1063/1.1568331} {\bibfield
  {journal} {\bibinfo  {journal} {J.\ Chem.\ Phys.}\ }\textbf {\bibinfo
  {volume} {118}},\ \bibinfo {pages} {8699--8705} (\bibinfo {year}
  {2003})}\BibitemShut {NoStop}%
\bibitem [{\citenamefont {Nielsen}\ \emph {et~al.}(2011)\citenamefont
  {Nielsen}, \citenamefont {Simesen}, \citenamefont {Bisgaard}, \citenamefont
  {Stapelfeldt}, \citenamefont {Filsinger}, \citenamefont {Friedrich},
  \citenamefont {Meijer},\ and\ \citenamefont
  {K{\"u}pper}}]{Nielsen:PCCP13:18971}%
  \BibitemOpen
  \bibfield  {author} {\bibinfo {author} {\bibfnamefont {J.~H.}\ \bibnamefont
  {Nielsen}}, \bibinfo {author} {\bibfnamefont {P.}~\bibnamefont {Simesen}},
  \bibinfo {author} {\bibfnamefont {C.~Z.}\ \bibnamefont {Bisgaard}}, \bibinfo
  {author} {\bibfnamefont {H.}~\bibnamefont {Stapelfeldt}}, \bibinfo {author}
  {\bibfnamefont {F.}~\bibnamefont {Filsinger}}, \bibinfo {author}
  {\bibfnamefont {B.}~\bibnamefont {Friedrich}}, \bibinfo {author}
  {\bibfnamefont {G.}~\bibnamefont {Meijer}}, \ and\ \bibinfo {author}
  {\bibfnamefont {J.}~\bibnamefont {K{\"u}pper}},\ }\bibfield  {title}
  {\enquote {\bibinfo {title} {Stark-selected beam of ground-state {OCS}
  molecules characterized by revivals of impulsive alignment},}\ }\href
  {\doibase 10.1039/c1cp21143a} {\bibfield  {journal} {\bibinfo  {journal}
  {Phys.\ Chem.\ Chem.\ Phys.}\ }\textbf {\bibinfo {volume} {13}},\ \bibinfo
  {pages} {18971--18975} (\bibinfo {year} {2011})},\ \Eprint
  {http://arxiv.org/abs/1105.2413} {arXiv:1105.2413 [physics]} \BibitemShut
  {NoStop}%
\end{thebibliography}%
\onecolumngrid
\end{document}